\documentclass[final,1p,times]{elsarticle}

\usepackage{graphicx}

\begin{document}

\title{The Basis of the Second Law of Thermodynamics in Quantum Field Theory}

\author{D.W. Snoke\footnote{Email: snoke@pitt.edu, Phone: 412-624-9007, Fax: 412-624-9163}}
\author{G. Liu}
\address{Department of Physics and Astronomy, University of Pittsburgh,
Pittsburgh, PA 15260, USA}
\author{S. Girvin}
\address{Department of Physics, Yale University,
New Haven, CT 06520, USA}

\begin{abstract}

The result that closed systems evolve toward equilibrium is derived entirely on the basis of quantum field theory for a model system, without invoking any of the common extra-mathematical notions of particle trajectories, collapse of the wave function, measurement, or intrinsically stochastic processes. The equivalent of the Stosszahlansatz of classical statistical mechanics is found, and has important differences from the classical version. A novel result of the calculation is that interacting many-body systems in the infinite volume limit evolve toward diagonal states (states with loss of all phase information) on the the time scale of the interaction time. The connection with the onset of off-diagonal phase coherence in Bose condensates is discussed. 
\end{abstract}
%\date{}

\maketitle
%\tableofcontents
% \renewcommand{\thesection}{\arabic{section}}
 
%  \renewcommand{\subsection}{\@startsection{subsection}{2}{0in}{.3 \baselineskip}{.3\baselineskip}{\it \normalfont}}

%\@startsection{<type>}{<level>}{<indent>}{<beforeskip>}{<afterskip>}{<style>}

\section{Introduction}

The controversy over the formulation of the Second Law of thermodynamics in terms of statistical mechanics of particles is well known \cite{ehrenbook}.  Boltzmann presented his H-theorem as a way of connecting the deterministic, time-reversible dynamics of particles to the already well known, time-irreversible Second Law. 
This theorem, in Boltzmann's formulation using classical mechanics, says that a many-particle system evolves irreversibly toward equilibrium via collision processes of the particles which can be calculated using known collision kinetics. Many scientists in his day found his arguments unpersuasive, however, partly because they involved notions of coarse graining which were ill-defined, and partly because the outcome of irreversibility seemed to contradict well-established ideas of time-reversibility such as Poincare cycles.

Later on, statistical mechanics was given firmer footing on the basis of quantum mechanical states, and this approach is used in most modern textbooks \cite{reif}. Typically, the notion of entropy is introduced in terms of counting states, and the equilibrium distributions of quantum particles are deduced on the basis of particle statistics. %but the $H$-theorem is not derived from fundamental quantum mechanical principles.  
The basic approach is the same as Boltzmann's, using probability and statistics of particles, with the only difference being that the particles have well-defined quantum states. This gives the impression that the notion of indivisible particles is crucial for understanding statistical mechanics. 
 
In this article, we show that by using modern quantum field theory it is possible to deduce irreversible behavior for a closed system entirely on the basis of wave mechanics without ever invoking the concept of a random ensemble of particles. Starting with a nonequilibrium many-body wave function, an equivalent of the Boltzmann scattering integral can be written down, by which expectation values of the wave function describing the system can be evolved deterministically and irreversibly toward equilibrium in the thermodynamic limit.  This method has already been used to successfully explain various real experiments, but we examine in detail the assumptions which are involved in justifying this approach. 

While one may interpret the many-body wave function as giving the probability of observation of discrete particles, it is not necessary to do so in order to calculate any of the values we calculate here; the equilibration follows entirely from the mathematical structure of the continuous wave function. This calculation supports the philosophical view that the field of quantum field theory is ``real.'' By this we mean that the notion of particles is a subordinate concept of the field theory, useful in some circumstances and dispensable in others.  In the case studied here of equilibration, the particle concept is dispensable. In general, particles are entirely deducible from the field theory as eigenstates (resonances) of the field. The many-body wave function is therefore a more fundamental physical description of the system.

A surprising result of the calculation here is that systems generally evolve toward diagonal states (many-body states with no phase coherence between different single-particle quantum states) on time scales comparable to the collision time.  Thus, the calculation shows a type of ``collapse'' of the wavefunction that comes entirely from the mathematics of quantum field theory without need for invoking external observers or other, more esoteric concepts.

Several other works have addressed the evolution of a quantum mechanical system toward equilibrium, notably Refs.~\cite{walls} and \cite{linden}. These calculations have typically taken the approach of using a ``bath'' or ``environment'' which interacts with a smaller subsystem; the bath is assumed to already be in equilibrium.  In the present work, we treat the entire closed system as one system, with the same equations of evolution for all parts, and all of which can be far from equilibrium. The essential puzzle we face is that in this case, quantum evolution is unitary and hence the von Neumann entropy of a closed system ought to be a constant of the motion.  This situation has been studied by Gellman and Hartle \cite{gell}, who used coarse-graining ideas to study the time evolution of quantum systems and the cross-over to classical irreversible evolution.  Polkovnikov \cite{polk} has introduced the concept of the microscopic ``diagonal entropy,'' which is the von Neumann entropy computed ignoring the contributions from the off-diagonal parts of the density matrix.  He has shown that the diagonal entropy has all the properties expected of the thermodynamic entropy.  We will present here a derivation of the quantum Boltzmann equation which yields irreversible time evolution and show that the quantum coherences represented by simple off-diagonal elements of the density matrix do indeed (generically) decay towards zero. As shown in Appendix A, an odd result of the Polkovnikov definition of diagonal entropy is that the entropy of a boson system near condensation decreases for some period of time. A generalized H-theorem can still be derived with a revision of the definition of entropy that also involves only diagonal elements, but which has less obvious connection to the von Neumann entropy. 

A key concept in modern quantum information theory that helps us resolve the paradox described above is that of entanglement entropy \cite{entangle,entangle2}.  A strange feature of quantum mechanics is that a system can be in a pure quantum state with zero entropy even though each of its subsystems has finite entropy.  This is because the entanglement between the different subsystems gives non-classical correlations which make a negative contribution to the entropy of the overall system.  For example, the reduced density matrix for a portion of a quantum system in its ground state obtained by tracing over the other parts of the system describes a mixed state with finite energy fluctuations and finite entropy \cite{entangle2}.  Coarse graining, in which we trace over short distance degrees of freedom to obtain a long-wavelength description \cite{gell}, can thus lead to the appearance of decoherence, classicality, and entropy.

We discuss below how apparently irreversible decoherence naturally arises within the approximations made in deriving the quantum Boltzmann equation. A new result here is that the average values of the off-diagonal elements of the generalized density matrix, which hold the information about the entanglement, all tend exponentially quickly toward zero. In the low-density limit, in which both bosons and fermions recover classical statistics, the ratio of the off-diagonal elements to the diagonal elements strictly vanishes. Thus, while the total von Neumann entropy does stay constant, the entropy in the off-diagonal terms becomes spread over so many possible entanglements that it becomes inaccessible to experiments. 

\section{Derivation of the Quantum Boltzmann Equation}

The equation which describes the evolution of a many-body system toward equilibrium is commonly known as the quantum Boltzmann equation; to be accurate, the term ``quantum Boltzmann equation'' should be used for an equation which describes both the real-space and momentum-space evolution of a many-body system, while the term ``Fokker-Planck equation'' should be used for an equation for the evolution in momentum space only, but in the case of a homogeneous gas, these are the same, and the two terms are often used interchangeably.  The full classical Boltzmann equation is usually assumed to include terms for the spatial drift of the particles (see, e.g., Ref.~\cite{snokebook}, Section 5.9). Therefore we will use the term ``classical Boltzmann integral'' when referring below to the analogous classical calculation for the evolution of the particle distribution function.

%A derivation for the quantum Boltzmann equation has been presented in Ref.~\onlinecite{snokebook}, Section 4.8, but here we will do a more rigorous derivation for the most general possible initial state. 

A basic result of quantum field theory which we will use here is that a quantum field can be written in terms of  second-quantization operators which obey the commutation relation 
\begin{equation}
[a^{ }_k,a^{\dagger}_k] = a^{ }_ka^{\dagger}_k - a^{\dagger}_ka^{ }_k = 1
\label{bosecomm}
\end{equation}
 for bosons, and the anticommutation relation 
 \begin{equation}
 \{a_k,a^{\dagger}_k\} = a_ka^{\dagger}_k + a^{\dagger}_ka_k = 1
 \end{equation}
for fermions. In a homogeneous system, the subscript corresponds to a wave vector $\vec{k}$. (We drop the vector notation on subscripts everywhere in this paper.) The mathematics of these operators goes back to Dirac \cite{dirac} and is presented in many recent textbooks such as Refs.~\cite{mandl} and \cite{baym}. These operators naturally lead to the language of particles, because the operator $a_k^{\dagger}$ 
acting on a wave state increases the eigenvalue of the operator $\hat{N}_k = a_k^{\dagger}a^{ }_k$ by 1 and the operator $a^{ }_k$ decreases the eigenvalue of $\hat{N}_k$ by 1, making it natural to treat $\hat{N}_k$ as giving an integer number of particles, and $a_k^{\dagger}$ as the ``creation'' operator and $a_k^{ }$ as the ``destruction'' operator for single particles.  More fundamentally, the $a_k^{\dagger}$ and $a_k^{ }$ operators represent the complex amplitude of a wave of the quantum field in mode $\vec{k}$, and $\hat{N}_k$ represents its intensity. We will often use the particle language in this paper, however, for convenience. Instead of saying ``expectation value of the operator $\hat{N}_k$'' we can just as well say ``average number of particles in state $\vec{k}$;'' instead of ``low field amplitude'' we can equivalently say ``low particle density,'' and instead of ``nonlinear mode-coupling terms in the wave equation'' we can refer to ``particle-particle interaction terms.'' %It is understood that in our context, ``particle'' is shorthand for ``eigenstate of the system in the limit of zero nonlinear terms.'' 

\renewcommand{\thesubsection}{\arabic{section}.\arabic{subsection}}

\subsection{Quantum Boltzmann Evolution for a General State}
\label{sect.qboltz}

A quantum field in the limit of zero nonlinear terms, i.e., zero particle-particle interactions, has eigenstates known as Fock states, or ``number'' states, which are defined by starting with a vacuum state and repeatedly using creation operators, as follows:
\begin{equation}
| \psi_i \rangle = \prod_{{{k}} }
\frac{\left(a^{\dagger}_{{k}}\right)^{N_{{k}}}}{\sqrt{N_{{k}}!}}|
\mbox{0}
\rangle.
\label{bath}
\end{equation}
Here $N_{{k}}$ is an integer which gives the occupation number of each state $\vec{k}$. (For fermion operators, we must apply the restriction $N_k \le 1$.) The ``vacuum'' state $|\mbox{0}\rangle$ is the zero-particle state, which means the ground state of the system, with no excitations. The set of all Fock states makes up a complete set of states, which means that any quantum state $|\psi\rangle$ can be written as a superposition of Fock states:
\begin{equation}
|\psi\rangle = \sum_n \alpha_n |n\rangle,
\label{focksum}
\end{equation}
where the states $|n\rangle$ are Fock states and $\alpha_n$ is the phase factor for each state. 

In a non-interacting system, the total Hamiltonian is given by $ H_0 = \sum \hat{N}_k \hbar\omega_k + E_0$, where $E_0$ is the energy of the ground state. %where the $\hat{N}_k$ with a hat refers to the number operator $\hat{N}_k = a^{\dagger}_ka^{ }_k$.
If there were no additional terms in the Hamiltonian, eigenstates $|n\rangle$ would keep the same number $N_k$ of quanta in each state $\vec{k}$ for all time, and there would be no change of the expectation value of the number operators. Interactions which allow the system to approach equilibrium arise from nonlinear terms which multiply creation and destruction operators acting on different $k$-states. Such terms arise from anharmonic terms in the Hamiltonian---the non-interacting Hamiltonian is equivalent to each $\vec{k}$-mode of the system being a perfect harmonic oscillator.

There are all manner of interactions; here, we assume a number-conserving interaction term which corresponds to the collision of two particles, i.e., applying destruction operators to two states and creation operators to two other states:
\begin{equation}
{\hat{V}} = \frac{1}{2}\sum_{{{k}}_1, {{k}}_2, {{k}}_3 } U_{{k}_1, {k}_2,
{k}_3, {k}_4} a^{\dagger}_{{k}_4} a^{\dagger}_{{k}_3}a^{ }_{{k}_2}a^{ }_{{k}_1}.
\label{int}
\end{equation}
In this sum, $\vec{k}_4$ is fixed by momentum conservation, $\vec{k}_1+\vec{k_2} = \vec{k}_3 + \vec{k}_4$, and $U_{{k}_1, {k}_2,{k}_3, {k}_4}$ is the associated energy of the transition; it is assumed to be symmetric on exchange of ${\vec{k}}_1$ with ${\vec{k}}_3$ or ${\vec{k}}_2$ with ${\vec{k}}_4$. A derivation of this type of term from wave functions in the infinite-volume limit can be found in Ref.~\cite{snokebook}, Section 5.5.

This four-operator interaction term may seem ponderous, but is actually the simplest interaction that allows equilibration in a closed system; an elastic scattering term with only two operators does not allow the particles to change energy, while a three-operator interaction does not conserve the number of particles, and therefore requires keeping track of at least two terms, one of which leads to creation of new particles and one of which leads to absorption of existing particles. 

%In all the following, we will assume a three-dimensional system, but we will drop the vector notation for the momenta $\vec{k}$ in the subscripts, for simplicity.   

The evolution of the system is found using time-dependent perturbation theory (e.g., Ref.~\cite{baym}). If the initial state of the system is
$|\psi_i\rangle$, and the state of the system at some later time
$t$ is  $|\psi_t\rangle$, the change in the average number of particles in state $\vec{k}$ is
given by
\begin{eqnarray}
d\langle \hat{N}_{{k}}\rangle &=& \displaystyle \langle \psi_t | \hat{N}_{{k}} | \psi_t \rangle -
\langle \psi_i | \hat{N}_{{k}} | \psi_i\rangle \label{comm}\\
\nonumber\\ 
&=&\displaystyle \langle \psi(t) |
e^{iH_0t/\hbar} \hat{N}_{{k}}e^{-iH_0t/\hbar} | \psi(t)
\rangle - \langle \psi_i | \hat{N}_{{k}} | \psi_i\rangle \nonumber\\
\nonumber\\ 
&=& \displaystyle\langle \psi_i | e^{(i/\hbar)\int {\hat{V}}(t) dt}  \hat{N}_{{k}} e^{-(i/\hbar)\int {\hat{V}}(t)
dt}| \psi_i \rangle \nonumber\\
&& \hspace{1cm} - \langle \psi_i | \hat{N}_{{k}} | \psi_i\rangle
\nonumber\\
\nonumber\\ 
&=& \displaystyle\langle \psi_i | e^{(i/\hbar)\int {\hat{V}}(t) dt} {[} \hat{N}_{{k}}, e^{-(i/\hbar)\int
{\hat{V}}(t) dt} {]} | \psi_i \rangle , \nonumber
\end{eqnarray}
where $|\psi(t)\rangle = e^{iH_0t/\hbar}|\psi_t\rangle$ and $\hat{V}(t) = e^{iH_0t/\hbar}\hat{V}e^{-iH_0t/\hbar}$ in the standard interaction notation. The operator $\hat{N}_{{k}}$ commutes with $H_0$, by definition. If it commutes with ${\hat{V}}$, there is no change in the occupation numbers over time. 

The exponential of the interaction operator can be written out as the series expansion,
%\begin{widetext}
\begin{eqnarray}
d\langle \hat{N}_{{k}}\rangle
 & = & \langle \psi_i | \left( 1 - (1/i\hbar) \int_{0}^t {\hat{V}}(t')dt' 
+  ... \right) \label{series} \\
&& \times \left( (1/i\hbar) \int_{0}^t dt'{[ } \hat{N}_{{k}} , {\hat{V}}(t') {]}  
+  %\right. \nonumber\\
%&&  \left. 
(1/i\hbar)^2 \int_{0}^t dt' \int_{0}^{t'} dt'' {[} \hat{N}_{{k}}
,{\hat{V}}(t'){\hat{V}}(t'') {]}+ ...
\right) | \psi_i \rangle  . 
\nonumber
\end{eqnarray}
%\end{widetext}
Let us look first at the lowest-order term of this expansion, which is
\begin{eqnarray}
d\langle \hat{N}_{{k}}\rangle
 & = & \frac{1}{i\hbar} \int_{0}^t dt'   \langle \psi_i | [  \hat{N}_{{k}} , {\hat{V}}(t') ]| \psi_i \rangle\nonumber\\
% &=& \frac{1}{i\hbar} \int_{0}^t dt'   \langle \psi_i | [  \hat{N}_{{k}} , {\hat{V}} ]| \psi_i \rangle \nonumber\\
 &=&  \frac{t}{i\hbar} \langle \psi_i | [  \hat{N}_{{k}} , {\hat{V}} ]| \psi_i \rangle.
\end{eqnarray}

For the particular case of two-body interactions given by (\ref{int}), we can
resolve the commutator by using %the relations
\begin{eqnarray}
{[} \hat{N}_{{k}}, a_{{k}'} {]} = -a_{{k}} \delta_{{{k}},{{k}'}}, \hspace{.3cm} {[} \hat{N}_{{k}}, a^{\dagger}_{{k}'} {]} = a^{\dagger}_{{k}}
\delta_{{{k}},{{k}'}},\label{universal}
\end{eqnarray}
which are valid for both boson and fermion creation and
destruction operators. We find
\begin{eqnarray}
&&{[} \hat{N}_{{k}}, {\hat{V}} {]} 
= 
\label{comm2} \\
&&\hspace{.5cm} \frac{1}{2}\sum_{{{k}}_1, {{k}}_2}(U_{D} \pm
U_{E})\left(a^{\dagger}_{k} a^{\dagger}_{k_3}a_{k_2}a_{k_1} -
a^{\dagger}_{k_1} a^{\dagger}_{k_2}a_{k_3}a_{k} \right) , \nonumber
\end{eqnarray}
where $\vec{k_3} = \vec{k}_1+\vec{k}_2 - \vec{k}$, and $U_{D}$ refers to the direct term and $U_{E}$ to the exchange term, and
the + sign is for bosons and the $-$ sign is for fermions. 

The lowest-order term of the expansion is therefore 
\begin{eqnarray}
d\langle \hat{N}_{{k}}\rangle &=& \frac{t}{2i\hbar}\sum_{{k}_1, {k}_2}(U_{D} \pm
U_{E})  \left( \langle \psi_i |a^{\dagger}_{k} a^{\dagger}_{k_3}a_{k_2}a_{k_1} | \psi_i \rangle  -
 \langle \psi_i | a^{\dagger}_{k_1} a^{\dagger}_{k_2}a_{k_3}a_{k} | \psi_i \rangle  \right).
 \label{firstorder}
\end{eqnarray}
If $|\psi_i\rangle$ is a Fock state, this term vanishes exactly; if the two creation operators restore the same two states that the destruction operators removed, then the operator has the form $N_kN_{k_1}$ and is equal for both terms, which cancel. More generally, if $|\psi_i\rangle$ is not a Fock state, (\ref{firstorder}) depends on the value of ``off-diagonal'' elements of the form $ \langle \psi_i |a^{\dagger}_{k} a^{\dagger}_{k_3}a_{k_2}a_{k_1} | \psi_i \rangle$. We will argue in Section \ref{sect.dephase} that these terms are generally negligible. 
%xx note that first order growth of these is prop. to i, therefore i's cancel and term will be real.

We therefore move on to the second-order terms of the expansion (\ref{series}). The first of these is
%\begin{widetext}
\begin{eqnarray}
d\langle \hat{N}_{{k}}\rangle & = & \frac{1}{\hbar^2}  \int_{0}^t 
dt' \int_0^t dt''
 \langle \psi_i |  {\hat{V}}(t') [  \hat{N}_{{k}} , {\hat{V}}(t'') ] | \psi_i \rangle \nonumber\\
% &=&  \frac{1}{\hbar^2}  \int_{0}^t  dt' \int_0^t dt'' \left( \langle \psi_i |  {\hat{V}}(t') \hat{N}_{{k}} {\hat{V}}(t'') | \psi_i  \rangle  -  \langle \psi_i |  {\hat{V}}(t') {\hat{V}}(t'')\hat{N}_{{k}}  | \psi_i  \rangle \right) \nonumber\\
 %&=&  \frac{1}{\hbar^2}  \int_{0}^t dt' \int_0^t dt'' \left( \langle \psi_i |  e^{iH_0t'/\hbar}{\hat{V}}e^{-iH_0t'/\hbar} \hat{N}_{{k}} e^{iH_0t''/\hbar}{\hat{V}}e^{-iH_0t''/\hbar}| \psi_i  \rangle \right. \nonumber\\
%&& \left. -  \langle \psi_i |  e^{iH_0t'/\hbar}{\hat{V}}e^{-iH_0(t'-t'')/\hbar} {\hat{V}}e^{-iH_0t''/\hbar}\hat{N}_{{k}}  | \psi_i  \rangle \right)\nonumber\\
% &=&  \frac{1}{\hbar^2}  \int_{0}^t  dt' \int_0^t dt'' \left( \langle \psi_i |  e^{iH_0t'/\hbar}{\hat{V}}e^{-iH_0(t'-t'')/\hbar} \hat{N}_{{k}} {\hat{V}}e^{-iH_0t''/\hbar}| \psi_i  \rangle \right. \nonumber\\
%&& \left. -  \langle \psi_i |  e^{iH_0t'/\hbar}{\hat{V}}e^{-iH_0(t'-t'')/\hbar} {\hat{V}}\hat{N}_{{k}} e^{-iH_0t''/\hbar} | \psi_i  \rangle \right)\nonumber\\
 &=&  \frac{1}{\hbar^2}  \int_{0}^t dt' \int_0^t dt'' 
 \langle \psi_i |  e^{iH_0t'/\hbar}{\hat{V}}e^{-iH_0(t'-t'')/\hbar} [\hat{N}_{{k}}, {\hat{V}}]e^{-iH_0t''/\hbar}| \psi_i  \rangle .
\end{eqnarray}
%\end{widetext}
We insert the general definition (\ref{focksum}) of the state $|\psi_i\rangle$ in terms of Fock states, and insert a sum over the projection operators for all Fock states, $\displaystyle\sum_m | m\rangle\langle m | = 1$, to obtain:
%\begin{widetext}
\begin{eqnarray}
d\langle \hat{N}_{{k}}\rangle & = &\sum_{m,n,n'} \frac{1}{\hbar^2}  \int_{0}^t dt' \int_0^t dt'' \  \alpha_{n'}^*\alpha_{n}
 \langle n' |  e^{iH_0t'/\hbar}{\hat{V}}e^{-iH_0(t'-t'')/\hbar} |m\rangle\langle m|[\hat{N}_{{k}}, {\hat{V}}]e^{-iH_0t''/\hbar}| n  \rangle  \nonumber\\
 &=& \sum_{m,n,n'} \frac{1}{\hbar^2}  \int_{0}^t dt' \int_0^t dt'' \  \alpha_{n'}^*\alpha_{n}
e^{i(E_{n'}-E_m)t'/\hbar} e^{-i(E_{n}-E_m)t''/\hbar} \langle n' |  {\hat{V}} |m\rangle\langle m|[\hat{N}_{{k}}, {\hat{V}}]| n  \rangle .\nonumber\\ \label{2ndorder1}
\end{eqnarray}
%\end{widetext}
Since $\hat{V}$ now acts just on Fock states, we can resolve:% the last factor as
%\begin{widetext}
\begin{eqnarray}
&&\sum_m  \langle n' |  {\hat{V}} |m\rangle\langle m|[\hat{N}_{{k}}, {\hat{V}}]| n  \rangle =  \nonumber
%  \frac{1}{2}\sum_{k_1', k_2'}(U_{D} + U_{E}) \left( \langle n |  {\hat{V}} |n'^{+}\rangle\sqrt{N_{k_1'}N_{k_2'}(1+N_{k_3'})(1+N_{k})} \right. \nonumber\\
 %&& \left. - \langle n |  {\hat{V}} |n'^-\rangle\sqrt{N_kN_{k_3'}(1+N_{k_2'})(1+N_{k_1'}) }  \right)\nonumber\\
% &=&  \frac{1}{2}\sum_{k_1', k_2'}(U_{D} + U_{E})(U_D'+U_E')   \nonumber\\
%&&\times \left( \langle n^+  |n'^{+}\rangle \sqrt{N_{k_1}N_{k_2}(1+N_{k_3})(1+N_{k})} \sqrt{N_{k_1'}N_{k_2'}(1+N_{k_3'})(1+N_{k})} \right. \nonumber\\
% && \left. - \langle n^- | n'^-\rangle \sqrt{N_kN_{k_3}(1+N_{k_2})(1+N_{k_1}) }
% \sqrt{N_kN_{k_3'}(1+N_{k_2'})(1+N_{k_1'}) }  \right) \nonumber
%&&\hspace{1cm} 
\frac{1}{2}\sum_{{k}_1', {k}_2',{k}_3'} 2(U_D\pm U_E) \sqrt{N_{k_1'}N_{k_2'}(1\pm N_{k_3'})(1\pm N_{k_4'})}   \\
&& \hspace{.5cm}\times \frac{1}{2}\sum_{{k}_1, {k}_2}(U_{D} \pm U_{E}) %\nonumber\\
%&&\hspace{1cm} \times
 \biggl( \sqrt{N_{k_1}N_{k_2}(1\pm N_{k_3})(1\pm N_{k})} 
  -  \sqrt{N_kN_{k_3}(1\pm N_{k_2})(1\pm N_{k_1})}\biggr) \langle n' | n'' \rangle. \nonumber \\  \label{step1}
\end{eqnarray}
%\end{widetext}
where $|n''\rangle$ is the same as state $|n\rangle$ but with 8 states changed by the 8 creation and destruction operators. The numbers $N_{k}$, $N_{k_1}$, etc., (without hats) are the occupation numbers of the $k$-states in the Fock state $|n\rangle$. 
The sum over states $|m\rangle$ is eliminated by the requirement that $|m\rangle$ be equal to $|n'\rangle$ with 4 states changed. The factor $\langle n' | n'' \rangle = \delta_{n',n''}$ eliminates the sum over states $n'$. Here we have assumed that all four $k$'s for the transition from $|n\rangle$ to $| m\rangle$ are different; if some $k$'s are the same there will be correction terms. However, momentum and energy conservation in the case of $|n\rangle = |n''\rangle$, discussed below, will make this impossible except in the case of forward scattering (no change of any momenta), in which case such terms make no contribution to change of the momentum distribution.  

If the eight changes from $|n\rangle$ to $|n''\rangle$ restore all the particle states back to their original states, so that $|n''\rangle = |n\rangle$, then this formula is simplified considerably. This forces $\vec{k}_1'$ and $\vec{k}_2'$ to equal $\vec{k}_1$ and $\vec{k}_2$ or vice versa (giving a factor of 2), or $\vec{k}_1'$ and $\vec{k}_2'$ to equal $\vec{k}_3$ and $k$ or vice versa. In the same way, $\vec{k}_{3'}$ can equal either $\vec{k}$ or $\vec{k}_3$ (giving the direct and exchange terms) if $\vec{k}_1'$ and $\vec{k}_2'$ equal $\vec{k}_1$ and $\vec{k}_2$, or $\vec{k}_{3'}$ can equal either $\vec{k}_1$ or $\vec{k}_2$ if $\vec{k}_1'$ and $\vec{k}_2'$ equal $\vec{k}_3$ and $\vec{k}$.  This leaves us with
%\begin{widetext}
\begin{eqnarray}
d\langle \hat{N}_{{k}}\rangle & = & \sum_{n}|\alpha_{n}|^2   \sum_{{k}_1, {k}_2} \frac{1}{2\hbar^2} \int_{0}^t dt' \int_0^t dt'' \   e^{i(E_{k_1}+E_{k_2} - E_{k_3} - E_k)t'/\hbar} e^{-i(E_{k_1}+E_{k_2} - E_{k_3} - E_k)t''/\hbar} \nonumber\\
&&\times
(U_{D} \pm U_{E})^2\biggl(  N_{k_1}N_{k_2}(1\pm N_{k_3})(1\pm N_{k}) 
  -  N_kN_{k_3}(1\pm N_{k_2})(1\pm N_{k_1}) \biggr).
  \label{step2}
\end{eqnarray}
%\end{widetext}

%xx if two states change and restore, remaining 4 can be phase rho^(2)

The time-dependent factors can be resolved as
\begin{eqnarray}
&&   \left(\int_{0}^t dt' \ e^{i\omega t'}\right)
 \left(\int_0^t dt'' \ e^{-i\omega t''}\right)\nonumber\\
&&\hspace{.5cm}= \   
\left|\frac{e^{i\omega t}-1}{\omega}\right|^2 = \frac{\sin^2(\omega t/2)}{\omega^2}.
\label{fk1}
\end{eqnarray}
We now make two approximations. The first is that the energy states are close enough together that there is some range $d\omega$ over which the energy-dependent factors and the density of states can be treated as constant, so that we can treat the sum over $k$ states as an integral over energy. This will always be true in the infinite-volume limit, in which case the states of the system are a continuum.  The second approximation is to assume that the state $|\psi_i\rangle$ is changing slowing compared to the rate of oscillation of the phase factors. Both of these approximations are the standard ones for deriving Fermi's golden rule for transition rates, and can be called the Markovian limit; we will discuss the breakdown of this approximation below. In this case, the time-dependent factor can be written as:
\begin{eqnarray}
 \lim_{t\rightarrow \infty}   \frac{\sin^2(\omega t/2)}{\omega^2} &=&    \delta(\omega)2\pi t.
\label{identity}
\end{eqnarray}
Taking $t$ as a small time interval $dt$, we then have for (\ref{2ndorder1}):  
%\begin{widetext}
%\begin{eqnarray}
%d\langle \hat{N}_{{k}}\rangle & = & \frac{2\pi t}{\hbar}  \sum_n |\alpha_n|^2 \ \frac{1}{2} \sum_{{k}_1, {k}_2}(U_{D} \pm U_{E})^2  \biggl( N_{k_1'}N_{k_2'}(1\pm N_{k_3'})(1\pm N_{k})    -  N_kN_{k_3}(1\pm N_{k_2})(1\pm N_{k_1}) \biggr)\nonumber\\
%&&\times\delta(E_{k_1}+E_{k_2} - E_{k_3} - E_k).
%\end{eqnarray} 
%\end{widetext}
%\begin{widetext}
\begin{eqnarray}
%\fbox{$\displaystyle
%\begin{array}{rcl}
\displaystyle\frac{d}{dt} \langle \hat{N}_{{k}}\rangle& = &\displaystyle \frac{2\pi}{\hbar}   \frac{1}{2} \sum_{{k}_1, {k}_2}(U_{D} \pm U_{E})^2  \delta(E_{k_1}+E_{k_2} - E_{k_3} - E_k)  \label{qboltz} \\
&&\times
\biggl(\langle \psi_i|\hat{N}_{k_1}\hat{N}_{k_2}(1\pm\hat{N}_{k_3})(1\pm \hat{N}_{k})| \psi_i\rangle   - \langle \psi_i|\hat{N}_k\hat{N}_{k_3}(1\pm \hat{N}_{k_2})(1\pm \hat{N}_{k_1})| \psi_i\rangle\biggr)\nonumber
.
%\end{array}
%$}\\
\end{eqnarray}
%\end{widetext}
This is nearly the standard quantum Boltzmann equation, or Fokker-Planck equation, for two-body interactions. The only remaining step is to factor the expectation values to have products like $\langle \hat{N}_k \rangle \langle \hat{N}_{k'}\rangle$ instead of $\langle \hat{N}_k  \hat{N}_{k'}\rangle$ etc. We will address this factoring in the next section. 

% It says, essentially, that the evolution of the distribution function $\langle N_k\rangle$ is given by applying the standard quantum mechanical rule, Fermi's golden rule \cite{baym}, to every possible transition. 

In going from (\ref{step1}) to (\ref{step2}), we restricted ourselves to terms in the sums where the second interaction term completely undid the changes in the state $|n'\rangle$ made by the first interaction term. If we had not done that, for a given set of $k$'s we would have terms of the form 
\begin{eqnarray}
&& \sum_{n} \alpha_{n''}^*\alpha_n \left(  \sqrt{N_{k_1'}N_{k_2'}(1\pm N_{k_3'})(1\pm N_{k_4'})} 
%\right. \nonumber\\
%&&\hspace{.5cm}\times \left.
\sqrt{N_{k_1}N_{k_2}(1\pm N_{k_3})(1\pm N_{k})} \right)  \nonumber\\
&& \hspace{.5cm}=  \sum_{n,n'} \left( \langle n'| \alpha_{n'}^* \right)a^{\dagger}_{k_1'}a^{\dagger}_{k_2'}a_{k_3'}a_{k_4'}a^{\dagger}_{k}a^{\dagger}_{k_3}a_{k_2}a_{k_1} (\alpha_{n} |n  \rangle) \nonumber\\
&& \hspace{.5cm} = \langle \psi_i | a^{\dagger}_{k_1'}a^{\dagger}_{k_2'}a_{k_3'}a_{k_4'}a^{\dagger}_{k_1}a^{\dagger}_{k_2}a_{k_3}a_{k} |\psi_i\rangle,
\end{eqnarray}
where the eight operators are those which take $|n\rangle$ to $|n''\rangle$. This is another, fourth-order ``off-diagonal'' phase term like the four-operator, second-order off-diagonal term found in (\ref{firstorder}).  As with that term, we will argue in Section~\ref{sect.dephase} that these fourth-order off-diagonal terms are negligible.

Last, we must examine the other second-order term in (\ref{series}). This is
\begin{eqnarray}
&& - \frac{1}{\hbar^2} \int_{0}^t dt' \int_{0}^{t'} dt'' \langle \psi_i | [ \hat{N}_{{k}}
,{\hat{V}}(t'){\hat{V}}(t'') ] | \psi_i \rangle \nonumber \\
&& \hspace{1cm} = 
 - \frac{1}{\hbar^2} \int_{0}^t dt' \int_{0}^{t'} dt'' \sum_{n,n'} \alpha_{n'}^*\alpha_n \left(\langle n' |  \hat{N}_{{k}}{\hat{V}}(t'){\hat{V}}(t'')|n\rangle
 % \right.\nonumber\\
%&&\hspace{.5cm}\left. 
- \langle n'|{\hat{V}}(t'){\hat{V}}(t'')\hat{N}_{{k}}  | n \rangle\right)\nonumber\\
&&  \hspace{1cm} - \frac{1}{\hbar^2} \int_{0}^t dt' \int_{0}^{t'} dt'' \sum_{n} \alpha_{n''}^*\alpha_n (N_k''-N_k)
%\nonumber\\
%&&\hspace{2cm}\times
 \langle n''|{\hat{V}}(t'){\hat{V}}(t'')   | n \rangle.
\end{eqnarray}
If we consider terms in which the state $|n''\rangle$ equals $|n\rangle$, then this term vanishes, since $N_k''$, which is the value of $N_k$ in the state $|n''\rangle$, is then equal to $N_k$. If we consider other terms, then we must deal with the same fourth-order off diagonal terms discussed above, which we will argue in Section~\ref{sect.dephase} are negligible.

\subsection{Factoring the Number Averages}
\label{sect.factor}

Equation (\ref{qboltz}) gives the rate of change of any average occupation number $\langle \hat{N}_k\rangle$ in terms of instantaneous averages of the form $\langle \psi_i| \hat{N}_{k_1}\hat{N}_{k_2}|\psi_i\rangle$ and $\langle \psi_i| \hat{N}_{k_1}\hat{N}_{k_2}\hat{N}_{k_3}|\psi_i\rangle$. (The two terms with four number operators actually cancel out.) What we would like, however, is an equation in terms of only $\langle \psi_i| \hat{N}_{k_1}|\psi_i\rangle$, $\langle \psi_i| \hat{N}_{k_2}|\psi_i\rangle$, etc., so that we only need to keep track of the average occupation numbers, not any correlations between the numbers in different momentum states.  This requires us to justify using $\langle \psi_i| \hat{N}_{k_1}\hat{N}_{k_2}|\psi_i\rangle = \langle \psi_i| \hat{N}_{k_1}|\psi_i\rangle\langle \psi_i| \hat{N}_{k_2}|\psi_i\rangle$. 

Since the interactions lead to real-space correlations or anticorrelations, this factorization will not, in general, be exactly true, but can be taken as a good approximation in an incoherent system. One way to justify this is to look at the implications of reasonable assumptions about the spatial correlations. We write the density-density correlation function,
\begin{eqnarray}
G^{(2)}(\vec{r}) = \langle \hat\rho(\vec{r})\hat\rho(0)\rangle,
\end{eqnarray}
where 
\begin{equation}
\hat\rho(\vec{r}) = \psi^{\dagger}(\vec{r})\psi(\vec{r}) = \frac{1}{{V}}\sum_{k_1,k_2} e^{-i(\vec{k}_1-\vec{k}_2)\cdot\vec{r}}a^{\dagger}_{k_1}a^{ }_{k_2}
\end{equation}
is the local density operator in terms of the spatial field operators. Substituting in this definition, we have
\begin{eqnarray}
G^{(2)}(\vec{r})  = \frac{1}{V^2}\sum_{{k}_1,{k}_2,{k}_3,{k}_4}e^{-i(\vec{k}_1-\vec{k}_2)\cdot\vec{r}}\langle a^{\dagger}_{{k}_1}a^{ }_{{k}_2}a^{\dagger}_{{k}_3}a^{ }_{{k}_4}\rangle.\nonumber\\
\end{eqnarray}
As we did in the previous section, we will assume that ``off-diagonal'' terms $\langle a^{\dagger}_{{k}_1}a^{ }_{{k}_2}a^{\dagger}_{{k}_3}a^{ }_{{k}_4}\rangle$, with all four $\vec{k}$'s different, are nearly zero, so that only terms with two sets of $\vec{k}$'s the same, to make number operators, will contribute to this sum. This gives us
\begin{eqnarray}
G^{(2)}(\vec{r})  = \frac{1}{V^2}\sum_{{k}_1,{k}_2}\left(\pm e^{-i(\vec{k}_1-\vec{k}_2)\cdot\vec{r}}+1\right)\langle \hat{N}_{{k}_1}\hat{N}_{{k}_2}\rangle,
\end{eqnarray}
where the $+$ sign is for bosons and the $-$ sign is for fermions. 

If there is no correlation in the many-body wave function at separation $\vec{r}$, we expect that $G^{(2)}(\vec{r})$ will have the value 
\begin{eqnarray}
 \langle \psi^{\dagger}(\vec{r})\psi(\vec{r})\rangle \langle \psi^{\dagger}(0)\psi(0)\rangle   &=&
 % \nonumber\\
%&&\hspace{1.5cm} 
\frac{1}{V^2} \sum_{{k}_1,{k}_2} e^{-i(\vec{k}_1-\vec{k}_2)\cdot\vec{r} } \langle a^{\dagger}_{k_1}a^{ }_{k_2}\rangle  \sum_{{k}_3,{k}_4} \langle a^{\dagger}_{k_3}a^{ }_{k_4} \rangle \nonumber\\
&=&  
\frac{1}{V^2} \sum_{{k}_1} \langle \hat{N}_{k_1}\rangle  \sum_{{k}_3} \langle \hat{N}_{k_3} \rangle \nonumber\\
&=& \bar{\rho}^2,
\end{eqnarray}
where $\bar{\rho} = N/V$ is the average number density, and we again assume vanishing off-diagonal terms which make only the number operators relevant. 
Integrating over all space, we obtain
\begin{eqnarray}
\int d^3r  \left(G^{(2)}(\vec{r})-\bar{\rho}^2\right) &=&  
 \pm\frac{1}{V}\sum_{{k}_1}\langle \hat{N}_{k_1}^2\rangle +\frac{1}{V}\sum_{{k}_1,{k}_2}\left(\langle \hat{N}_{k_1}\hat{N}_{k_2}\rangle-\langle \hat{N}_{k_1}\rangle\langle \hat{N}_{k_2}\rangle \right).\nonumber
\label{g2int}
\end{eqnarray}
The first sum on the right side can be taken as negligible compared to the second, since it consists of just the $\vec{k}_2 = \vec{k}_1$ terms of the first term in the second sum. 

Suppose that $G^{(2)}(\vec{r})$ deviates from $\bar{\rho}^2$ only for some limited range in space, e.g.,
\begin{equation}
G^{(2)}(\vec{r}) = \bar{\rho}^2(1-e^{-r^2/a^2}),
\end{equation}
where $a$ is some characteristic length. Then the integral over all space is
\begin{eqnarray}
\int d^3r  \left(G^{(2)}(\vec{r})-\bar{\rho}^2\right)  \sim -\bar{\rho}^2 a^3.  
\end{eqnarray}
Then we have
\begin{eqnarray}
 \frac{1}{V}\sum_{{k}_1,{k}_2}\left(\langle \hat{N}_{k_1}\hat{N}_{k_2}\rangle-\langle \hat{N}_{k_1}\rangle\langle \hat{N}_{k_2}\rangle \right) \sim -\bar{\rho}^2 a^3 ,\nonumber\\
\end{eqnarray}
or 
\begin{eqnarray}
\sum_{{k}_1,{k}_2}\left(\langle \hat{N}_{k_1}\hat{N}_{k_2}\rangle-\langle \hat{N}_{k_1}\rangle\langle \hat{N}_{k_2}\rangle \right) %\nonumber\\
%&& \hspace{1cm}
&\sim & - \frac{a^3}{V} \sum_{{k}_1,{k}_2} \langle \hat{N}_{k_1}\rangle\langle \hat{N}_{k_2}\rangle.\end{eqnarray}
The sum on the right-hand side is small compared to the second term in the sum on the left-hand side, as long as $a$ is not extensive with the whole volume. We therefore have 
\begin{eqnarray}
\sum_{{k}_1,{k}_2}\left(\langle \hat{N}_{k_1}\hat{N}_{k_2}\rangle-\langle \hat{N}_{k_1}\rangle\langle \hat{N}_{k_2}\rangle \right) \approx 0. 
\label{factordev}
\end{eqnarray} 
In other words, on average, the deviation of $\langle \hat{N}_{k_1}\hat{N}_{k_2}\rangle$ from $\langle \hat{N}_{k_1}\rangle\langle \hat{N}_{k_2}\rangle$ is zero.

We can make the same type of argument for the factorization of $\langle \hat{N}_{k_1}\hat{N}_{k_2}\hat{N}_{k_3}\rangle$ using higher-order correlation functions. (The terms with four $\hat{N}_k$ operators in (\ref{qboltz}) cancel.) In general, a localized correlation or anticorrelation in real space will lead to no significant $k$-space number correlations. The case of a Bose-Einstein condensate, which has infinite-range correlations, requires re-examination of the factoring of the number averages. We discuss some aspects of Bose-Einstein condensation below, in Section~\ref{sect.dephase2}.

\section{Solution of the Quantum Boltzmann Equation}

We now have a tractable equation for numerical simulations. We can simply define the occupation number $\langle \hat{N}_k\rangle$ for all $\vec{k}$ and calculate the rate of change for each $\langle\hat{N}_k\rangle$ at every point in time. In this section we examine the numerical solution method in detail.

In all of the above, we have assumed that the $k$-states are defined for a set of discrete $\vec{k}$ values.  In the limit of infinite volume, however, the spacing between states $\vec{k}$ disappears, and the distribution function $\hat{N}_{k}$ is a continuous function defined for all $\vec{k}$. In that case, we can set $\langle \hat{N}_k\rangle$ equal to its average value in a small element of phase space $d^3k$. This allows us to convert the double summation to a double integral, 
\begin{eqnarray}
\frac{d\langle\hat{N}_k\rangle}{dt} &=&
\frac{2\pi}{\hbar}\left(\frac{V}{(2
\pi)^3}\right)^2 \ \frac{1}{2}\int d^3k_1 \ d^3k_2  \  |U_{D} \pm U_{E}|^2 \delta(E_{{k}_1}+E_{{{k}}_2} - E_{{{k}}}-E_{{{k'}}})\nonumber\\
&&\times\left[\langle \hat{N}_{{k}_1}\rangle \langle \hat{N}_{k_2}\rangle (1 \pm \langle \hat{N}_{k}\rangle)(1
\pm \langle \hat{N}_{k'}\rangle) %\right.\nonumber\\&&\left.
%\hspace{.5cm}
-  \langle \hat{N}_{k}\rangle \langle \hat{N}_{k'}\rangle(1 \pm \langle
\hat{N}_{k_1}\rangle)(1 \pm \langle \hat{N}_{k_2}\rangle)\right],
%&&\hspace{.7cm}\times,
\label{avgboltz}
\end{eqnarray}
where $V$ is the volume, and the $+$ signs are for bosons and the $-$ signs for fermions. If we further assume that the system is isotropic,  i.e. the distribution function depends only on the magnitude of ${\vec{k}}$ and not on the direction, then the angles in the integral (\ref{avgboltz}) can be eliminated by analytical integration, to reduce this integral to just a double integral over two energies, which can then be computed numerically. We can then write the distribution function simply as $N(E_k) = \langle\hat{N}_k\rangle$.  At a given point in time, the change $dN(E)/dt $ can be computed for all $E$, and then each $N(E)$ is
updated according to 
\begin{equation}
N(E) \rightarrow N(E) +  \frac{dN(E)}{dt} dt,
\label{update}
\end{equation}
where $dt$ is some small time interval, chosen such that the change of $N(E)$ is small during any one update.

\begin{figure}[h]
\begin{center}
 \includegraphics[width=2.4in]{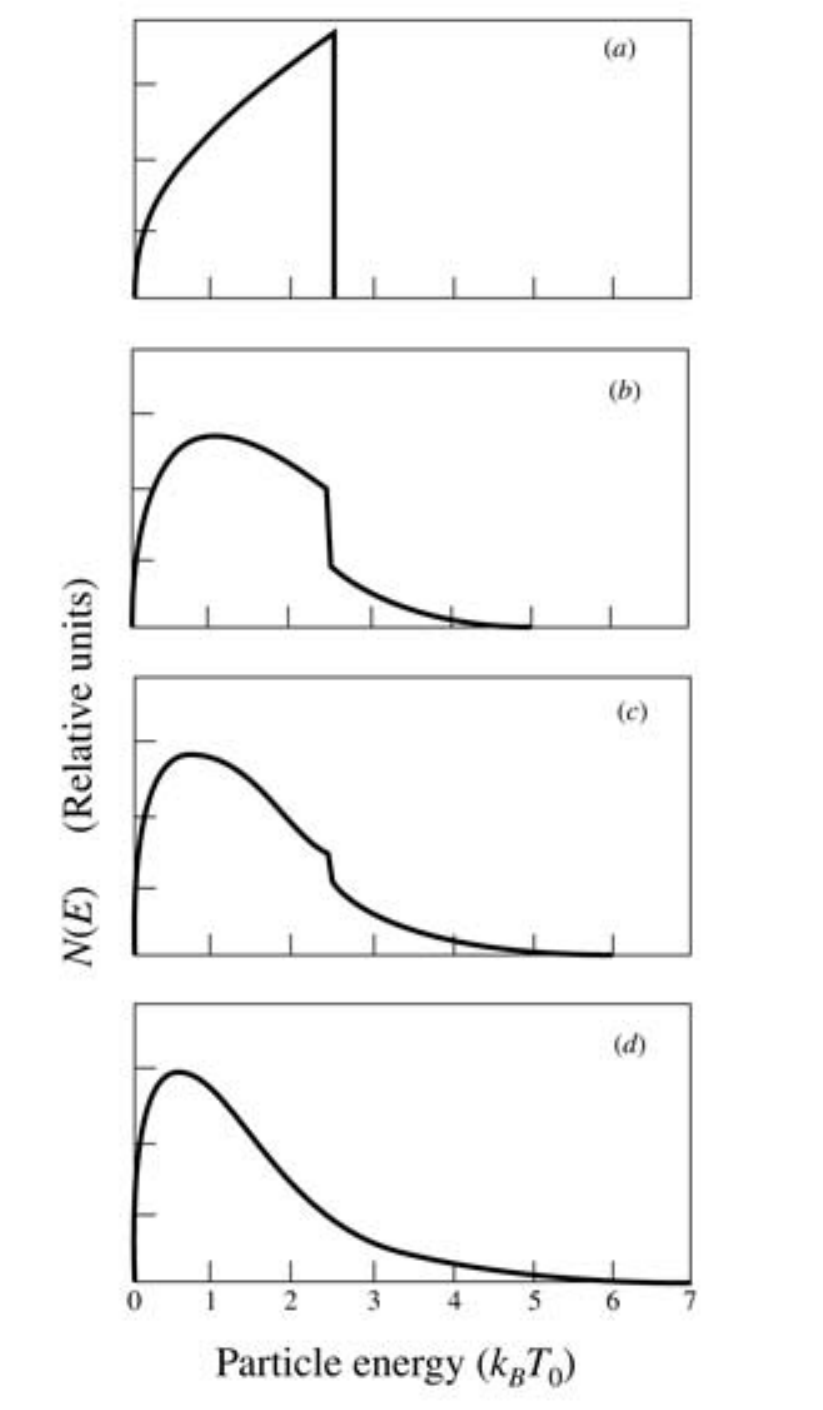} 
\end{center}
\caption{Nonequilibrium evolution of an isotropic gas of bosons at low density interacting with the potential (\ref{int}), found by numerically solving Eq.~(\ref{avgboltz}). The initial state (a) corresponds to equal occupation of all states up to a cutoff ($N(E)$ is proportional to the density of states ${\cal D}(E) \propto \sqrt{E}$). The distribution (b) is after each particle has undergone one scattering event, on average, i.e., the total amount of change of the distribution $N(E)$ is equal to the total integral of $N(E)$.  The distribution (c) is after two scattering events per particle. The distribution in (d), after four scattering events per particle, is less than 1\% different from a Maxwell-Boltzmann distribution. The horizontal energy axis is given in units of $k_BT_0$, where $T_0$ is the
final temperature. From Ref.~\protect\cite{snoke1989}.}
\label{fig.bosevolve}
\end{figure}

In other words, we have a prescription for determining the approach of a nonequilibrium system toward equilibrium {\em deterministically}. Figure \ref{fig.bosevolve} gives the solution generated by a computer for bosons at low density starting from a nonequilibrium distribution. After five times the characteristic scattering time $\tau$, the distribution does not change significantly---the system has come to equilibrium. This is not a Monte Carlo simulation. In this calculation, no random numbers were used, and no individual particles were tracked. The distribution function $N(E)$ was evolved deterministically using the isotropic quantum Boltzmann equation, such that each successive state was completely determined by the previous state. 

We can easily determine the equilibrium distribution by finding the steady-state solutions of the quantum Boltzmann equation. In equilibrium, we must have $d{N}(E_k)/dt =0$
for all $\vec{k}$. This will occur if for all the different scattering processes, the forward and backward rates for any process are the same:
\begin{eqnarray}
&&N(E_{k_3})N(E_{k_2})(1 \pm N(E_{k_1}))(1
\pm N(E_k)) \label{balance}\\
&&-  N(E_k)N(E_{k_1})(1 \pm N(E_{k_2})(1 \pm N(E_{k_3}) = 0.\nonumber
\end{eqnarray}
This is the principle of detailed balance. It can easily be verified that a distribution of the form
\begin{equation}
N(E) = \frac{1}{e^{\alpha + \beta E }\mp 1}
\label{balancesol}
\end{equation}
satisfies this condition, where the $-$ sign is for bosons and the + sign for fermions. The constants $\alpha$
and
$\beta$ are determined by the conditions
\begin{equation}
N = \sum_k N_k = \int N(E){\cal D}(E) dE, \hspace{.5cm}
\label{nconser}
%\end{equation}
%\begin{equation}
U = \sum_k E_k N_k = \int EN(E){\cal D}(E) dE,
\end{equation}
where ${\cal D}(E)$ is the density of $k$-states at energy $E$.  These conditions follow from number and energy conservation, which follow from the fact that the total number operator $N$ and the total energy operator $U$ commute with the Hamiltonian.
We can find $\alpha$ and $\beta$ in terms of standard thermodynamic quantities by noting that if $\alpha \gg 1$, the
equilibrium distribution is equal to $N(E) = e^{-\alpha}e^{-\beta E}$. Equating this to the standard equilibrium
distribution $N(E) = e^{\mu/k_BT}e^{-E/k_BT}$, which is called the Maxwell-Boltzmann distribution, we then have
$\beta = 1/k_BT$ and
$\alpha = -\mu/k_BT$, or
\begin{equation}
N(E) = \frac{1}{e^{(E-\mu)/k_BT}\mp 1},
\end{equation}
which is the well-known equilibrium distribution for quantum particles; when the sign is negative it is called the Bose-Einstein distribution and when the sign is positive it is called the Fermi-Dirac distribution. The quantum Boltzmann equation is therefore another way of deducing the standard equilibrium distributions. 

The Maxwell-Boltzmann limit $\mu \ll E_0$ corresponds to low particle density both the Fermi-Dirac and the Bose-Einstein distributions. The equilibrium solution shown in Figure~\ref{fig.bosevolve} for the distribution of the particles at late times converges to the Maxwell-Boltzmann distribution times the density of states of the particles, which for the three-dimensional gas in the model is proportional to $\sqrt{E}$ (see, e.g., Ref.~\cite{snokebook}, Section 1.8). %Note that Boltzmann's constant $k_B$ does not appear in (\ref{qboltz}), however. The energy distribution at late times is determined entirely by the contraints of energy and number conservation. The constant $k_B$ just gives the constant of proportionality between the energy per particle and the temperature scale we have chosen.

It is easy to show that the equilibrium solution of the quantum Boltzmann equation is the same
even if we invoke some process other than the two-body elastic scattering mechanism assumed here. For example, suppose we have an electron-photon interaction of the form
\begin{equation}
{\hat{V}} = \sum   (a^{\dagger}_{p} + a_{p}) a^{\dagger}_{e2}a_{e1},
\end{equation}
where $a_{e1}$ and $a_{e2}$ are fermion destruction operators for electrons in states 1 and 2, respectively, and $a_p$ is the operator for a photon which couples the two states. This gives the detailed balance equation
\begin{equation}
N_{e1}(1-N_{e2})N_p - N_{e2}(1-N_{e1})(1+N_p) = 0.
\end{equation}
If the electrons are in a Fermi-Dirac equilibrium distribution, then the ratio of the electron population terms is
\begin{equation}
\frac{N_{e2}(1-N_{e1})}{N_{e1}(1-N_{e2})} = e^{-(E_2-E_1)/k_BT},
\end{equation}
which then implies
\begin{equation}
N_p = \frac{1}{e^{(E_2-E_1)/k_BT}-1},
\end{equation}
which is the Planck distribution, equal to the Bose-Einstein equilibrium distribution with $\alpha = 0$. 
Therefore, just as with any other equilibrium distribution, we can also derive the Planck distribution entirely from the field theory. Although the Planck distribution historically played a major role in convincing people of the particle nature of waves, we see here that it fundamentally arises because of the field operator commutation relation (\ref{bosecomm}), which led to the final states factor $(1+N_k)$ in (\ref{qboltz}).  The commutation relation, in turn, follows from the wave nature of the quantum field, without reference to any particle collapse or measurement (see Ref.~\cite{snokebook}, Sections 4.1-4.3). 

\begin{figure}
\begin{center}
 \includegraphics[width=2.1in]{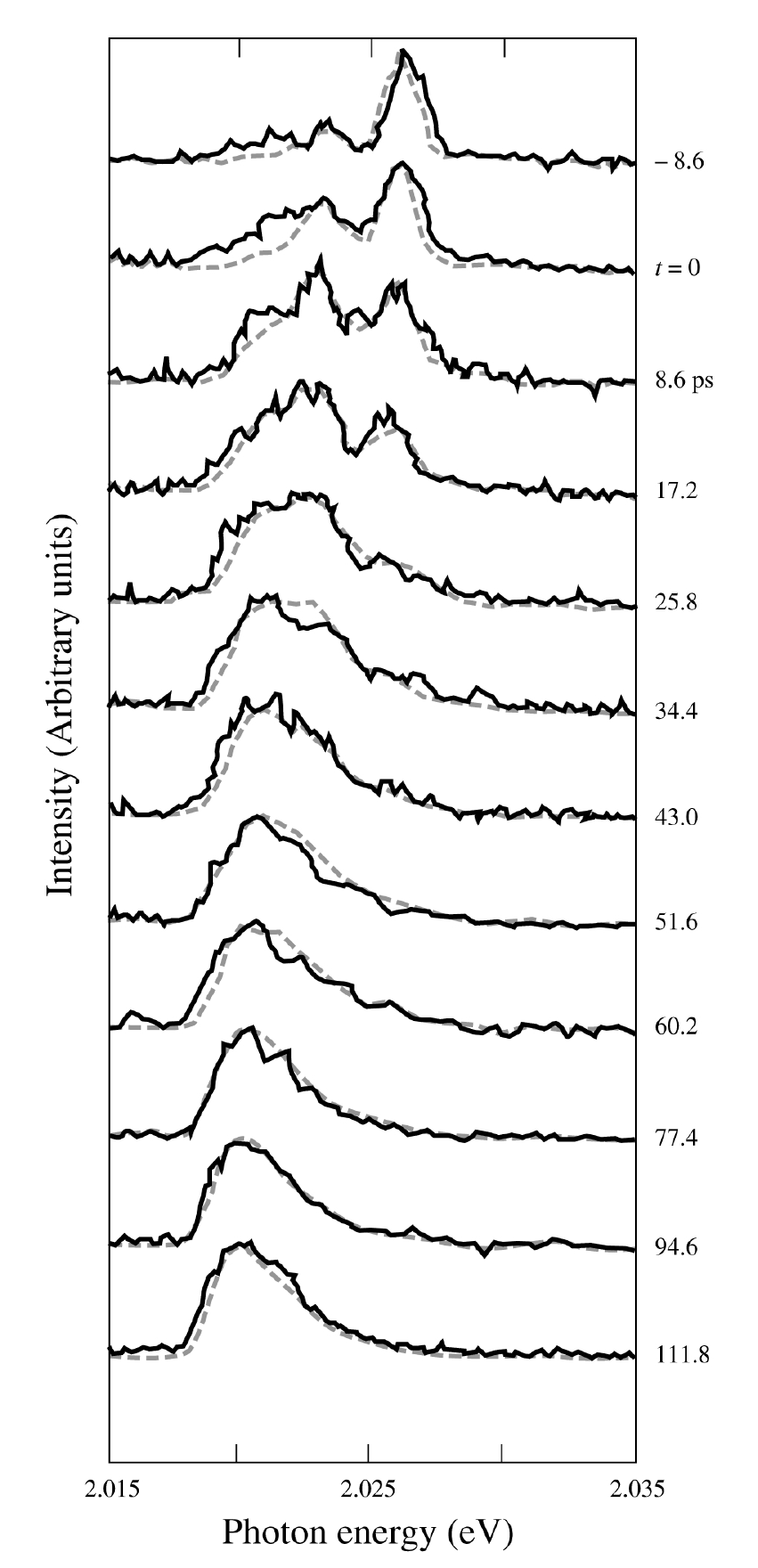} 
\end{center}
\caption{Solid lines: energy distribution of excitons in the semiconductor Cu$_2$O measured at various times following a laser pulse with temporal width 2 ps and maximum intensity at time $t=0$. Dashed lines: solution of the quantum Boltzmann equation for the time evolution of the population using a deformation-potential matrix elemennt for exciton-phonon scattering. The theory gives a Maxwell-Boltzmann distribution at all late times. (From Ref.~\protect\cite{sbc}).}
\label{fig.exc}
\end{figure}

The nonequilibrium evolution predicted by the quantum Boltzmann equation has been verified experimentally \cite{adp}.   Figure \ref{fig.exc} shows an example of the evolution of a population of excitons in a semiconductor. Modern ultrafast optics experiments can resolve the energy distribution of these particles on time scales short compared to the time to equilibrate. Therefore we can observe the approach to equilibrium and compare it to the evolution given by the quantum Boltzmann equation. As seen in this figure, there is a very good fit of the experiment to the theory. Only one fit parameter was used to fit the data at all times, namely, the matrix element for the strength of the interaction. The initial distribution was determined by the laser pulse which created the nonequilibrium distribution. Long after the laser pulse, the particles remain in a Maxwell-Boltzmann distribution.

\section{Where Did Irreversibility Enter In?}
\label{sect.irr}

The universal evolution toward equilibrium seen in the numerical solutions of the quantum Boltzmann equation is equivalent to the content of the H-theorem of Boltzmann, which says that when entropy is calculated for a nonequilibrium system, it always increases until the system reaches equilibrium. Appendix A shows the connection of the quantum Boltzmann equation to the H-theorem in terms of various definitions of entropy. As seen in the previous section, though, the basic result of irreversible equilibration does not depend on any particular definition of entropy.

But in a closed, energy-conserving system, we expect the dynamics will be time-reversible.  In other words, if start with state $| i \rangle$ and evolve to a state $| j\rangle$ at some later time, then we expect that if we started with the system in state $| j\rangle$, and reverse all the momenta, we should get back to state $| i \rangle$. 

The results of the previous sections appear to contradict this view. We began with a closed, energy-conserving system, namely a quantum mechanical field Hamiltonian with no interactions with any external system, and deduced the expectation values of the many-body wave function as it evolves deterministically toward equilibrium according to the proper wave equation; we never invoked collapse, measurement, observation, or randomness. From beginning to end we have treated only the wave function, without invoking particles at all except to identify them as the natural energy eigenstates of the system. Yet we get an irreversible, deterministic approach to equilibrium. %from our quantum wave solutions---if we reverse the momenta $\vec{k}$ in the distribution function at any time it will still evolve toward equilibrium, not away from it. 
How have we gone from a time-reversible system to one with apparent irreversibility?

Of course, along the way we have made several approximations in order to generate the quantum Boltzmann equation. Is it possible that these approximations are responsible for the irreversibility? Let us now examine in detail the assumptions
which were made. 

Let us first address an apparent inconsistency which may already have occured to the reader. On one hand, in using (\ref{identity}), we assumed that the time interval for the evolution was long enough that the $t \rightarrow \infty$ limit could be used, while in writing (\ref{qboltz}) we assumed that $t$ is short. Both of these limits can be valid. We pick $dt$ short enough that $N(E_k)$ of any given state is almost unchanged, but also long compared to a cycle of the phase factors which are relevant in that energy range.

This natural energy range is determined by the variation of the occupation number $N(E_k)$ with energy. 
If  $N(E_k)$ is a continuous function of $k$, then there will always be some energy range $(E_k, E_k+dE)$ over which $N(E_k)$ can be considered constant.  This defines the ``natural'' energy range $\Delta E$, which in turn defines a natural frequency $\omega = \hbar/\Delta E$.  For the function $\sin^2(\omega t/2)/(2\pi\omega^2 t)$ to be well approximated by the delta-function $\delta(\omega)$ as we did in (\ref{identity}),
we therefore want to choose $dt$ such that
\begin{equation}
\left(\frac{1}{N(E_k)}\frac{\partial N(E_k)}{\partial t}\right)^{-1} \ll  dt \ll \frac{\hbar}{\Delta E}.
\end{equation}

It is possible to have physical situations, however, in which the rate $\partial N(E_k)/\partial t$ is large, forcing the time interval $dt$ to be small, while $N(E_k)$ is strongly varying over a narrow range of energy, so that the natural energy range $\Delta E$ is small, and therefore the period $1/\omega$ is large, such that there is no range of time which satisfies both assumptions above. In this case one cannot use the quantum Boltzmann equation. Instead, one must use a non-Markovian evolution equation with a quantum memory kernel, in which memory is included of the state of the system at earlier times. Such methods, generally known as ``quantum kinetics,'' have been developed at length by Haug and
Koch \cite{haug-koch}. The general effect of using quantum kinetics is to introduce oscillations in the response of the system on very short time scales, typically of the order of femtoseconds for solid state systems. The solutions always approach an equilibrium distribution irreversibly on long time scales, however. This is because there will always be some scattering into regions of phase space where $N(E_k)$ is slowly varying enough for the quantum Boltzmann limit to apply.  

Another approximation which we used in deriving the quantum Boltzmann equation was restricting the perturbation theory to the leading order in the series expansion (\ref{series}). Would including higher-order terms remove the irreversibility? No, these higher-order terms just give multi-particle interactions instead of two-body interactions. Higher-order expansions of Fermi's golden rule are well known (see, e.g. Ref.~\cite{snokebook}, Section~8.1) and just give higher-order corrections to the matrix element used in the transition rates, i.e. corrections to the scattering cross section beyond the Born approximation.

Some may object that the quantum Boltzmann equation only keeps track of the average value of $N_k$ and does not keep track of fluctuations. There are two reasons why accounting for fluctuations will not restore reversibility. First, as discussed in Section~\ref{sect.factor}, we expect the numbers of particles in different $k$-states to be uncorrelated, which means that in an element of phase space $d^3k$ which includes many $k$-states, the number fluctuations should be governed by Poisson statistics for uncorrelated systems. In this case the number fluctuations are proportional to $\sqrt{N(E_k)}$, and therefore the ratio of fluctuations to the total number decreases as $1/\sqrt{N(E_k)}$, making them negligible in any macroscopic system. Furthermore, fluctuations do not lead to periodic behavior. The mean distribution $N(E_k)$ will still obey the evolution given by the quantum Boltzmann equation.

Where did the irreversibility enter in, then? The place to look is the iteration procedure (\ref{update}).  Note that the quantum Boltzmann equation (\ref{qboltz}) gives the change in the average occupation number $N_k$, using only the information of the average occupation numbers for the states. In the derivation of the quantum Boltzmann equation, we threw out terms of form 
\begin{equation}
\rho^{(2)}_{k_1,k_2,k_3,k} \equiv \langle \psi_i | a^{\dagger}_{k_1}a^{\dagger}_{k_2}a_{k_3}a_{k} |\psi_i\rangle, 
\end{equation}
and
\begin{equation}
\rho^{(4)}_{k_1,k_2,k_3,k_4,k_5,k_6,k_7,k} \equiv \langle \psi_i | a^{\dagger}_{k_7}a^{\dagger}_{k_6}a_{k_5}a_{k_4}a^{\dagger}_{k_3}a^{\dagger}_{k_2}a_{k_1}a_{k} |\psi_i\rangle. \nonumber\\
\end{equation}
One way of speaking of this is to say that we keep only the ``diagonal'' information of the state of the system, and discard the ``off-diagonal'' information. This terminology is a generalization of the density matrix formalism, where the first-order density matrix of a quantum system is defined as (see Ref.~\cite{snokebook}, Section 9.2)
\begin{eqnarray}
\rho = \left( 
\begin{array}{cccc}
\langle a^{\dagger}_{k_1}a^{ }_{k_1}\rangle &  \langle a^{\dagger}_{k_1}a^{ }_{k_2}\rangle & \langle a^{\dagger}_{k_1}a^{ }_{k_3}\rangle & \ldots\\ 
\langle a^{\dagger}_{k_2}a^{ }_{k_1}\rangle &  \langle a^{\dagger}_{k_2}a^{ }_{k_2}\rangle & \langle a^{\dagger}_{k_2}a^{ }_{k_3}\rangle & \ldots\\ 
\langle a^{\dagger}_{k_3}a^{ }_{k_1}\rangle &  \langle a^{\dagger}_{k_3}a^{ }_{k_2}\rangle & \langle a^{\dagger}_{k_3}a^{ }_{k_3}\rangle & \ldots\\ 
.\\
.\\
.
\end{array}
\right)
\label{singledens}
\end{eqnarray}
The diagonal elements of this matrix are the average occupation numbers, while the off-diagonal terms account for phase correlations in the system. We can generalize to define as ``off-diagonal'' any expectation values which do not involve only matched pairs of creation and destruction operators of the form $a^{\dagger}_ka_k$. 

Suppose that the system starts in a purely diagonal state, with all off-diagonal terms of all orders strictly equal to zero. After a short time $dt$, while the quantum Boltzmann equation gives the correct value for the change of $\langle \hat{N}_k\rangle$, the state of the system will have evolved to a different superposition of Fock states, and therefore the off-diagonal terms may be nonzero. The procedure (\ref{update}) amounts to replacing the state of the system with a purely diagonal state after every time step $dt$. 

Let us compare this to the standard derivation of the H-theorem in terms of classical particles. The classical Boltzmann integral looks just like the quantum Boltzmann equation (\ref{avgboltz}) derived above, in the limit when the particle density is low so that the $(1 \pm N_k)$ final states terms can be dropped.  Just as we have used a diagonal state with a continuous $\langle\hat{N}_k\rangle$ for the momentum distribution of the particles, the classical Boltzmann integral describes the state of the system in terms of an continuous velocity distribution function, which keeps track of the number of particles with a given velocity, but not the exact position and velocity of individual particles.   

Suppose that at time $t=0$ we have an isotropic, nonequilibrium velocity distribution of classical particles, and we evolve this distribution to a new velocity distribution at a later time time $t=dt$ via the classical Boltzmann integral. Following the same iterative approach described above for the evolution of a system using the quantum Boltzmann equation, we can then use the classical Boltzmann integral to evolve the distribution to the next time step, and so on until the system reaches equilibrium. This was Boltzmann's approach.  But since the microscopic laws are time-reversible, a problem arises.  If the velocities of all the particles at time $dt$ are reversed, then the system must evolve back to its original state at time $t=0$ and away from equilibrium. The classical Boltzmann integral will never show this behavior, however, because the overall velocity distribution for an isotropic system would be unchanged by reversing all the individual velocities. %Therefore Boltzmann integral cannot be valid for all cases, and the H-theorem is not strictly proven.

The resolution of this which gained acceptance in statistical mechanics is the {\em Stosszahlansatz} (for a review of this history, see Ref.~\cite{ehrenbook}). This assumption says that while some states of the system do exist which would give time-reversed behavior, for any given velocity distribution the number of such states is extremely small compared to the total number of states with the same average velocity distribution.  Effectively, then, at each time step in the iteration of the classical Boltzmann integral, we replace the exact velocity distribution of the particles with an average distribution. This loss of information ensures the irreversibility of the evolution \cite{vankamp}. In this approach, the H-theorem is thus seen as a statement of high probability, not absolute necessity. The chances of having a set of velocities which exactly reverse the time dynamics and move the system away from equilibrium is vanishingly small, but not strictly zero. 

The same loss of information happens in the numerical evolution of the quantum Boltzmann equation. Keeping track of all the phase factors of the superpositions of different Fock states in the full many-body quantum state is like keeping track of all the individual particle velocities in the classical case. If we did keep track of these terms, then we could write down the exact quantum mechanical state at time $dt$, and using this as an initial state, by reversing time we could recover the initial state at time $t=0$.  Replacing the state of the system with a diagonal state at each time step in the evolution of the quantum Boltzmann equation is like replacing the exact velocity distribution with an average velocity distribution at each time step when using the classical Boltzmann equation. The loss of the off-diagonal information, known as ``dephasing,'' ensures irreversibility.
\vspace{.2cm} 

\section{Calculation of the Off-Diagonal Time Evolution}
\label{sect.dephase}

As discussed above, the elimination of off-diagonal phase information plays the same role in the quantum calculation as the Stosszahlansatz in the classical calculation. The quantum calculation allows to us go one step further, however. Since we have a closed system with the full quantum mechanical many-body wave function written down, we can compute the dephasing rate. Then we can ask how valid an approximation it is to discard the phase information as we did in Section~\ref{sect.qboltz}. 
In that section, the evolution of $\langle \hat{N}_k\rangle$ depends, in principle, on the values of all of the different $\rho^{(2)}$ and $\rho^{(4)}$ terms. We note that the evolution of $\langle \hat{N}_k\rangle$ depends {\em only} on these terms, that is, on averages of the phase factors, and not on detailed knowledge of single values of the phase factors. %This means that if we can calculate the evolution of these two types of averages, then we know all we would need to know to reverse the evolution of the distribution of $\langle \hat{N}_k\rangle$ to make it move away from equilibrium. 

In the rest of this section, we show how to do this. We will focus on the evolution of the $\rho^{(2)}$ terms, and return to discuss the $\rho^{(4)}$ and other terms at the end of this section. We define the operator $\hat\rho^{(2)}_{k,k',k'',k'''} = a^{\dagger}_ka^{\dagger}_{k'}a^{ }_{k''}a^{ }_{k'''}$, and assume that the four momenta are all different, because the case when two or more momenta are the same will be greatly suppressed due to the requirement of momentum conservation in the interaction Hamiltonian which gives rise to evolution of the $\rho^{(2)}$ terms.

Following a similar approach to that of Section~\ref{sect.qboltz}, we write for a given off-diagonal term,
\begin{eqnarray}
 d\langle \hat\rho^{(2)}_{k,k',k'',k'''}\rangle &=& \
%&&\hspace{1cm}
 \displaystyle \langle \psi_t | \hat\rho^{(2)}_{k,k',k'',k'''} | \psi_t \rangle -
\langle \psi_i | \hat\rho^{(2)}_{k,k',k'',k'''} | \psi_i\rangle \nonumber\\
\nonumber\\ 
&=& \displaystyle \langle \psi(t) |
\hat\rho^{(2)}_{k,k',k'',k'''}(t) | \psi(t)
\rangle - \langle \psi_i | \hat\rho^{(2)}_{k,k',k'',k'''}| \psi_i\rangle .  
\end{eqnarray}
As we did for (\ref{series}), we expand the time-dependent exponential operators as
%\begin{widetext}
\begin{eqnarray}
d\langle \hat{\rho}^{(2)}_{{k,k',k'',k'''}}\rangle &=&\displaystyle \langle \psi_i |\left(1-\frac{1}{i\hbar}\int_0^t \hat{V}(t')dt' +\frac{1}{(i\hbar)^2}\int_0^t dt' \int_0^{t'} dt'' \hat{V}(t')\hat{V}(t'')+\ldots  
\right)
\hat{\rho}^{(2)}_{{k,k',k'',k'''}}(t) 
\nonumber \\
&&\times  \left(1+ \frac{1}{i\hbar}\int_0^t\hat{V}(t')dt' +\frac{1}{(i\hbar)^2}\int_0^t dt' \int_0^{t'} dt'' \hat{V}(t')\hat{V}(t'')+\ldots
\right) | \psi_i \rangle\nonumber\\
&&
 -\langle \psi_i | \hat{\rho}^{(2)}_{{k,k',k'',k'''}}| \psi_i\rangle .\label{rhoevol}
\end{eqnarray}
%\end{widetext}
The lowest-order term is
\begin{eqnarray}
 \langle \psi_i |(\hat{\rho}^{(2)}_{{k,k',k'',k'''}}(t)-\hat{\rho}^{(2)}_{{k,k',k'',k'''}} )| \psi_i\rangle ;
% &=& \sum_{n,n'} \alpha_{n'}^*\alpha_n \langle n'| (e^{iH_0t/\hbar} \hat{\rho}^{(2)}_{{k,k',k'',k'''}}e^{-iH_0t/\hbar}-\hat{\rho}^{(2)}_{{k,k',k'',k'''}} )|n\rangle,
\end{eqnarray}
the first-order terms are
\begin{equation}
 \displaystyle  \frac{1}{i\hbar}\int_0^t dt' \langle \psi_i | [\hat{\rho}^{(2)}_{{k,k',k'',k'''}}(t),\hat{V}(t')] 
 | \psi_i \rangle;
\end{equation}
while the second-order terms are
\begin{eqnarray}
&& \frac{1}{\hbar^2}\int_0^t dt' \int_0^t dt'' \ \langle\psi_i|\hat{V}(t') \hat{\rho}^{(2)}_{{k,k',k'',k'''}}(t)\hat{V}(t'') |\psi_i\rangle \nonumber\\
&& -\frac{1}{\hbar^2}\int_0^t dt' \int_0^{t'} dt'' \langle\psi_i| \hat{V}(t')\hat{V}(t'') \hat{\rho}^{(2)}_{{k,k',k'',k'''}}(t)|\psi_i\rangle\nonumber\\
&&-\frac{1}{\hbar^2} \int_0^t dt' \int_0^{t'} dt''  \langle\psi_i|\hat{\rho}^{(2)}_{{k,k',k'',k'''}}(t) \hat{V}(t')\hat{V}(t'')|\psi_i\rangle.\nonumber\\
\label{2ndorder}
\end{eqnarray}
We cannot assume that $\hat{\rho}^{(2)}_{{k,k',k'',k'''}}$ commutes with $H_0$, as we did for $N_k$ in the derivation of the quantum Boltzmann equation, so we must keep all of the $e^{iH_0t/\hbar}$ terms in the definitions of the time-dependent operators for the moment. In the following sections we compute each order of the expansion (\ref{rhoevol}) separately.

\subsection{Zero-Order Off-Diagonal Evolution}

As before, we write $|\psi_i\rangle$ as a sum of Fock states, to have the zero-order term
\begin{eqnarray}
%\langle \psi_i |(e^{iH_0t/\hbar} \hat{\rho}^{(2)}_{{k,k',k'',k'''}}e^{-iH_0t/\hbar}-\hat{\rho}^{(2)}_{{k,k',k'',k'''}} )| \psi_i\rangle
d\langle \hat\rho^{(2)}_{k,k',k'',k'''}\rangle &=&% \nonumber\\
%&&\hspace{.5cm} 
\sum_{n,n'} \alpha_{n'}^*\alpha_n 
\left( \langle n'| (e^{iH_0t/\hbar} \hat{\rho}^{(2)}_{{k,k',k'',k'''}}e^{-iH_0t/\hbar}|n\rangle 
%\right. \nonumber\\
%&&\hspace{.5cm}\left. 
- \langle n'|\hat{\rho}^{(2)}_{{k,k',k'',k'''}} )|n\rangle \right).\nonumber\\
\end{eqnarray}
Applying the time-varying exponential factors to the Fock states gives us 
\begin{eqnarray}
 d\langle \hat\rho^{(2)}_{k,k',k'',k'''}\rangle &=& 
%\nonumber\\
%&&\hspace{.5cm} 
\sum_{n,n'} \alpha_{n'}^*\alpha_n (e^{i(E_k+E_{k'}-E_{k''}-E_{k'''})t/\hbar}-1) %\nonumber\\
%&&\hspace{1cm}\times
\langle n'|\hat{\rho}^{(2)}_{{k,k',k'',k'''}}|n\rangle
\nonumber\\
&&\hspace{.5cm} \simeq  \frac{it}{\hbar}(E_k+E_{k'}-E_{k''}-E_{k'''})%\nonumber\\
%&&\hspace{1cm}\times 
\sum_{n,n'} \alpha_{n'}^*\alpha_n\langle n'|\hat{\rho}^{(2)}_{{k,k',k'',k'''}} |n\rangle \nonumber\\
&&\hspace{.5cm}=  \frac{it}{\hbar}(E_k+E_{k'}-E_{k''}-E_{k'''}) \langle \psi_i |\hat{\rho}^{(2)}_{{k,k',k'',k'''}}|\psi_i\rangle .\label{zerorot}
\end{eqnarray}
Thus, if there are nonzero $\hat\rho^{(2)}$ terms, these will rotate in phase proportional to the degree of violation of energy conservation.  

\subsection{First-Order Off-Diagonal Evolution}
\label{sect.dephase1}

We start by working on the first-order term. Inserting the definition of $\hat{V}$, we have
%\begin{widetext}
\begin{eqnarray}
&& \langle \psi_i | [\hat{\rho}^{(2)}_{{k,k',k'',k'''}}(t),\hat{V}(t')] | \psi_i \rangle  =\\
&&\hspace{1cm} \sum_{{k}_1,{k}_2,{k}_3} \frac{U}{2}  \langle \psi_i |[ e^{iH_0t}a^{\dagger}_k a^{\dagger}_{k'}a_{k''}a_{k'''}e^{-iH_0t},e^{iH_0t'}a^{\dagger}_{k_4} a^{\dagger}_{k_3} a^{ }_{k_2}a^{ }_{k_1} e^{-iH_0t'}]| \psi_i \rangle,\nonumber
\end{eqnarray} 
%\end{widetext}
where we abbreviate the momentum-dependent matrix element simply as $U$. As we did in deriving the quantum Boltzmann equation, we write the initial state as a superposition of Fock states of the form (\ref{focksum}):
%\begin{widetext}
\begin{eqnarray} 
&& \langle \psi_i | [\hat{\rho}^{(2)}_{{k,k',k'',k'''}}(t),\hat{V}(t')] | \psi_i \rangle = \nonumber\\
&&   \hspace{1cm} \sum_{n,n'} \alpha_{n'}^*\alpha_n \sum_{{k}_1,{k}_2,{k}_3}   \frac{U}{2}  ( \langle n' |e^{iH_0t/\hbar}a^{\dagger}_k a^{\dagger}_{k'}a_{k''}a_{k'''}e^{-iH_0t/\hbar} e^{iH_0t'/\hbar}a^{\dagger}_{k_1}  a^{\dagger}_{k_2} a_{k_3}a_{k_4} e^{-iH_0t'\hbar} |n \rangle \nonumber \\
&&\hspace{1.5cm} - \langle n' |e^{iH_0t'/\hbar}a^{\dagger}_{k_1} a^{\dagger}_{k_2} a_{k_3}a_{k_4} e^{-iH_0t'/\hbar}  e^{iH_0t/\hbar}a^{\dagger}_k a^{\dagger}_{k'}a_{k''}a_{k'''}e^{-iH_0t/\hbar} |n \rangle).
%&& \sum_{n,n'} \alpha_{n'}^*\alpha^{ }_n \sum_{k_1,k_2,k_3}   \frac{U}{2}  \left(e^{i(E_{k}+E_{k'}-E_{k''}-E_{k'''})(t-t')/\hbar}  \langle n' |a^{\dagger}_k a^{\dagger}_{k'}a_{k''}a_{k'''}   a^{\dagger}_{k_4}  a^{\dagger}_{k_3} a^{ }_{k_2}a^{ }_{k_1} |n \rangle  \right. \nonumber \\
 % &&\left. -e^{-i(E_{k}+E_{k'}-E_{k''}-E_{k'''})(t-t')/\hbar}   \langle n' |a^{\dagger}_{k_4} a^{\dagger}_{k_3} a^{ }_{k_2}a^{ }_{k_1}  a^{\dagger}_k a^{\dagger}_{k'}a_{k''}a_{k'''} |n \rangle \right).
  \label{firstexpans}
 \end{eqnarray} 
 For $n' = n$, this becomes
\begin{eqnarray}
   &&  \sum_{n} |\alpha^{ }_n|^2  2(U_D\pm U_E)  \left(e^{i(E_{k}+E_{k'}-E_{k''}-E_{k'''})(t-t')/\hbar} N_k^{(n)} N_{k'}^{(n)}(1\pm N_{k''}^{(n)})(1\pm N_{k'''}^{(n)}) \right.\nonumber \\
  && \left. \hspace{.5cm} -e^{-i(E_{k}+E_{k'}-E_{k''}-E_{k'''})(t-t')/\hbar}  N_{k''}^{(n)} N_{k'''}^{(n)}(1\pm N_{k}^{(n)})(1\pm N_{k'}^{(n)}) \right),
\end{eqnarray}
%\end{widetext}
where, as in Section~\ref{sect.qboltz}, $U_D$ and $U_E$ are the direct and exchange interaction constants, and the numbers $N_{k}^{(n)}$, $N_{k_1}^{(n)}$, etc., without hats, are the occupation numbers of the $k$-states in the Fock state $|n\rangle$. The $\pm$ signs here give the cases of bosons and fermions, as in the quantum Boltzmann equation.
If $n' \ne n$, we are left with off-diagonal terms, either $\hat\rho^{(2)}$ or higher-order phase terms, namely a six-operator $\hat\rho^{(3)}$ term or an eight-operator $\rho^{(4)}$ phase term. We will discuss these below.

Performing a time integral for the case $n'= n$, we have
\begin{eqnarray}
e^{i\omega t}\int_0^t dt' e^{-i\omega t'} &=&  %e^{i\omega t} \frac{e^{-i\omega t}-1}{-i\omega} \nonumber\\
 (1+i\omega t + \ldots) \frac{1-i\omega t+\ldots -1}{-i\omega}% \nonumber\\
%&\simeq& \frac{-i\omega t}{-i\omega} 
\ \simeq \ t .
\label{firsttime}
\end{eqnarray}

Thus, for a small time interval $t$, the first-order evolution from a diagonal state gives
%\begin{widetext}
\begin{eqnarray}
%\fbox{$\displaystyle
&&\frac{d}{dt} \langle \hat{\rho}^{(2)}_{k,k',k'',k'''}\rangle = \label{firstphase}\\
&&\hspace{.5cm} \frac{2i}{\hbar}(U_D\pm U_E)\langle \psi_i| \left(\hat{N}_{k''} \hat{N}_{k'''}(1\pm \hat{N}_{k})(1\pm \hat{N}_{k'})-\hat{N}_k \hat{N}_{k'}(1\pm \hat{N}_{k''})(1\pm \hat{N}_{k'''})|\psi_i\rangle\right).
%$}
\nonumber
\end{eqnarray}
%\end{widetext}
%(Actually the terms with $N_kN_{k'}N_{k''}N_{k'''}$ cancel out.)
This term gives us phase coherence that accumulates proportional to the net scattering rate connecting the four states, but without any condition of energy conservation. This term is imaginary, which multiplies the imaginary prefactor in (\ref{firstorder}), giving a real change of the expectation value $\langle \hat{N}_k\rangle$.  Since this term increases linearly with $t$, if the phase factors start out at zero value, the change in population in (\ref{firstorder}) will increase as $t^2$, which will be negligible on short time scales. But if the phase factors grow in time to have nonzero values, then they could contribute to the evolution of $\langle \hat{N}_k\rangle$.

As mentioned above, if we do not require $|n'\rangle = |n\rangle$ in equation (\ref{firstexpans}), we will have terms with higher-order off-diagonal terms instead of just number operators $\hat{N}_k$. In this case we will have terms like the following:
%\begin{widetext}
\begin{eqnarray}
 \langle n' | e^{iH_0t}a^{\dagger}_k a^{\dagger}_{k'}a^{ }_{k''}a^{ }_{k'''}e^{-iH_0t}
e^{iH_0t'}a^{\dagger}_{k_4} a^{\dagger}_{k_3} a^{ }_{k_2}a^{ }_{k_1} e^{-iH_0t'} | n \rangle   \nonumber\\   
-    \langle n' |e^{iH_0t'}a^{\dagger}_{k_4} a^{\dagger}_{k_3} a^{ }_{k_2}a^{ }_{k_1} e^{-iH_0t'}e^{iH_0t}a^{\dagger}_k a^{\dagger}_{k'}a_{k''}a_{k'''}e^{-iH_0t}
| n \rangle. \nonumber\\
\end{eqnarray}
%\end{widetext}
If none of the operators $a^{ }_{k_1},  a^{ }_{k_2}, a^{\dagger}_{k_3}$, or $a^{\dagger}_{k_4}$ act on the same $k$-states as the momenta in $\hat{\rho}^{(2)}_{{k,k',k'',k'''}}$, i.e., if all the operators in the interaction term commute with $\hat{\rho}^{(2)}_{{k,k',k'',k'''}}$, then these two terms will cancel. 

To deal with the case when the interaction term does not commute with $\hat{\rho}^{(2)}_{{k,k',k'',k'''}}$, we must deal with a number of possibilities. Appendix B treats these in detail. The general result is that if there are nonzero $\rho^{(2)}$ factors, then these can contribute to the first-order evolution of $\hat{\rho}^{(2)}_{{k,k',k'',k'''}}$. These other $\rho^{(2)}$ off-diagonal factors always appear in sums, such as 
\begin{equation}
\sum_{{k}_1,{k}_2} f(N_k,N_{k'},N_{k''},N_{k'''})\hat\rho^{(2)}_{k,k',k_1,k_2},
\end{equation}
where $f(N_k,N_{k'},N_{k''},N_{k'''})$ is some function of the occupation numbers, which can be either positive or negative. We might in general expect that these sums will average to zero, but as we will see below, even if they do not, there will be an overall evolution term to suppress these $\rho^{(2)}$ factors.

\subsection{Second-Order Off-Diagonal Evolution}
\label{sect.dephase2}

We now proceed to the second-order terms (\ref{2ndorder}).   First, we examine the time integrals. We assume that the operators in the first $\hat{V}(t')$ undo whatever changes have been done by $\hat{V}(t'')$; otherwise we will not be left with $\langle \psi_i| \hat{\rho}^{(2)}_{k,k',k'',k'''}|\psi_i\rangle$. Terms which do not do this will involve higher-order off-diagonal phase correlation terms $\hat\rho^{(3)}$, $\hat\rho^{(4)}$, etc.  These terms in principle can contribute to a second-order contribution to the evolution of $\langle \psi_i |\hat{\rho}^{(2)}| \psi_i \rangle$, but as with the first-order calculation, we take a perturbative approach that the $\hat\rho^{(2)}$ terms will dominate over higher-order off-diagonal terms. What we are interested in now is any terms proportional to $\langle \psi_i |\hat{\rho}^{(2)}| \psi_i \rangle$, so that we can see what happens to existing off-diagonal phase accumulation. 

If  $\hat{V}(t')$ undoes whatever changes have been done to $|\psi_i\rangle$ by $\hat{V}(t'')$, then each of the three terms in (\ref{2ndorder}) gives a factor $e^{i(E_{k_1}+E_{k_2}-E_{k_3}-E_{k_4})t'/\hbar}$ $\times e^{-i(E_{k_1}+E_{k_2}-E_{k_3}-E_{k_4})t''/\hbar}$, but in the first case $t'$ and $t''$ are integrated from 0 to $t$ separately, while in the last two terms $t'$ is integrated from 0 to $t$ but $t''$ is integrated from 0 to $t'$. The first case gives a factor $(2\pi t/\hbar)\delta(E_{k_1}+E_{k_2}-E_{k_3}-E_{k_4})$ as given in (\ref{identity}). The second case can be written as
\begin{eqnarray}
\int_0^t dt' \int_0^{t'}  dt'' \ e^{i\omega t'}e^{-i\omega t''} 
&=& \int_0^t dt' \int_0^{t} dt'' \ e^{i\omega t'}e^{-i\omega t''} - \int_0^t dt' \int_{t'}^{t} dt'' \ e^{i\omega t'}e^{-i\omega t''}\nonumber\\
%&=& \int_0^t dt' \int_0^{t} dt'' \ e^{i\omega t'}e^{-i\omega t''} + \int_0^t dt' \int_{t-t'}^{0} dt''' \ e^{i\omega t'}e^{-i\omega (t-t''')}\nonumber\\
&=& \int_0^t dt' \int_0^{t} dt'' \ e^{i\omega (t'-t'')}
- \int_0^t dt' \int_{0}^{t-t'} dt''' \ e^{-i\omega (t-t')}e^{i\omega t'''}\nonumber \\
&=& \int_0^t dt' \int_0^{t} dt'' \ e^{i\omega t'}e^{-i\omega t''} - \int_0^t dt'''' \int_{0}^{t''''} dt''' \ e^{-i\omega t''''}e^{i\omega t'''}.
\nonumber \\
\end{eqnarray}
The second term on the right in the last line is equal to the term on the left, and therefore we have
\begin{eqnarray}
\int_0^t dt' \int_0^{t'}  dt'' \ e^{i\omega t'}e^{-i\omega t''} %\nonumber\\
%&&\hspace{1cm}
&=& \frac{1}{2}\int_0^t dt' \int_0^{t} dt'' \ e^{i\omega t'}e^{-i\omega t''} \nonumber\\
&=& \frac{\pi t}{\hbar}\delta(E_{k_1}+E_{k_2}-E_{k_3}-E_{k_4}),\nonumber\\
\end{eqnarray}
i.e., ${1}/{2}$ of the first case. 

%$t'''=t-t'', t'' = t-t'''$ 
%$t''=t'-> t'''=t-t'$
%$t''=t-> t'''= 0$
%$t'''' = t-t'$

For convenience we write one term of the $\hat{V}$ sum as $\hat{V}_i = Ua^{\dagger}_{k_4}a^{\dagger}_{k_3} a_{k_2}a_{k_1}$. After removing the time dependence, the second order term of interest is
\begin{eqnarray} 
&\langle \psi_i |\left( \hat{V}_{i'}  \hat{\rho}^{(2)}\hat{V}_{i}-\frac{1}{2}\hat{V}_{i'}\hat{V}_{i}  \hat{\rho}^{(2)} -\frac{1}{2}\hat{\rho}^{(2)}\hat{V}_{i'}\hat{V}_{i} \right) | \psi_i \rangle  .
\label{2ndorderexp} 
\end{eqnarray}
This is the generalized Lindblad operator for a many-body system, used often in quantum optics \cite{qo}.
If $\hat{V}_i$ and $\hat{V}_{i'}$ both commute with $\rho^{(2)}$, then this vanishes. Thus, all of the terms in $\hat{V}$ which do not include a creation or destruction operator with at least one of the four momenta $k, k', k'', k'''$ in $\hat{\rho}^{(2)}$ do not contribute to any change of this dephasing term. This makes sense, since only interactions with these states should contribute to this dephasing term.

We must therefore consider each type of $\hat{V}_i$ term that does not commute with $\hat{\rho}^{(2)}_{k,k',k'',k'''}$. We will take these in order: terms with one momentum the same as in $\hat{\rho}^{(2)}_{k,k',k'',k'''}$, terms with two momenta the same as in $\hat{\rho}^{(2)}_{k,k',k'',k'''}$, terms with three the same, and terms with four the same. 

The interaction term $\hat{V}_i$ will not commute with $\hat{\rho}^{(2)}_{k,k',k'',k'''}$ if one of the four operators in $\hat{V}_i$ acts on one of the same momenta as the four in $\hat{\rho}^{(2)}_{k,k',k'',k'''}$.
We assume that $\hat{V}_{i'}$ has the same four operators as in $\hat{V}_i$, so that all the changes in $|\psi_i\rangle$ made by $\hat{V}_i$ are reversed, as discussed above. In this case, (\ref{2ndorderexp}) will have terms like the following:
\begin{eqnarray}
&&a_k^{\dagger}a_k^{\dagger}a_k^{ } - \frac{1}{2}a_k^{\dagger}a_k^{ }a_k^{\dagger}- \frac{1}{2}a_k^{\dagger}a_k^{\dagger}a_k^{ }
% \\ &=& \frac{1}{2}a_k^{\dagger}a_k^{\dagger}a_k^{ } -\frac{1}{2}a_k^{\dagger}(a_k^{\dagger}a_k^{ }+1) \\
= - \frac{1}{2}a^{\dagger}_k,
\label{2ndcomm}
\end{eqnarray}
which is true for both bosons and fermions, using the fermion property that a double creation operator vanishes. The remaining operator in (\ref{2ndcomm}) goes back into the definition of $\rho^{(2)}$, leaving behind three operators in each of $\hat{V}_i$ and $\hat{V}_{i'}$.

Setting $\vec{k}_1 = \vec{k}$ in $\hat{V}_i$, we can pick the momenta in $\hat{V}_{i'}$  four different ways to restore the same Fock state:
\begin{eqnarray}
&& a^{\dagger}_{k}a^{\dagger}_{k_2} a^{ }_{k_3}a^{ }_{k_4}a^{\dagger}_{k_4} a^{\dagger}_{k_3} a^{ }_{k_2} a^{ }_{k}
+ a^{\dagger}_{k}a^{\dagger}_{k_2} a^{ }_{k_4}a^{ }_{k_3}a^{\dagger}_{k_4} a^{\dagger}_{k_3} a^{ }_{k_2}  a^{ }_{k}
 \nonumber\\
&& +a^{\dagger}_{k_2}a^{\dagger}_k a^{ }_{k_3}a^{ }_{k_4}a^{\dagger}_{k_4} a^{\dagger}_{k_3} a^{ }_{k_2} a^{ }_{k}
+ a^{\dagger}_{k_2}a^{\dagger}_k a^{ }_{k_4}a^{ }_{k_3}a^{\dagger}_{k_4} a^{\dagger}_{k_3} a^{ }_{k_2}a^{ }_{k}.\nonumber\\
\end{eqnarray}
If $U$ is a constant, this just gives a factor of 4  multiplying the $(U/2)^2$ prefactor that comes from the two $\hat{V}$ terms in second order; more generally it gives the prefactor $\frac{1}{2}U_D(U_D \pm U_E)$. There will be another four terms giving $\pm\frac{1}{2}U_E(U_D \pm U_E)$ if we take $\vec{k}_2 = \vec{k}$, thus giving $(U_D \pm U_E)^2$, which is the same matrix element as in the quantum Boltzmann equation (\ref{qboltz}). 

Setting $\vec{k}_3$ or $\vec{k}_4$ in $\hat{V}_i$ equal to $k$ gives two more sets of four terms, using 
\begin{eqnarray}
&&a_k^{ }a_k^{\dagger}a_k^{\dagger} - \frac{1}{2}a_k^{ }a_k^{\dagger}a_k^{\dagger}- \frac{1}{2}a_k^{\dagger}a_k^{ }a_k^{\dagger} = \pm \frac{1}{2}a^{\dagger}_k.
\end{eqnarray}
where the $+$ is for bosons and the $-$ is for fermions.
%xx watch out-- also pick up - signs for every interchange

 Thus, for the momentum $k$ in $\hat{\rho}^{(2)}_{{k,k',k'',k'''}}$,  we have a factor of the form
\begin{eqnarray}
&& \frac{1}{2}(U_D \pm U_E)^2
%\\
%&&\hspace{.5cm}\times
\left[\pm \hat{N}_{k_1}\hat{N}_{k_2}(1\pm \hat{N}_{k_3})- \hat{N}_{k_2}(1\pm \hat{N}_{k_3})(1\pm \hat{N}_{k_4})\right].  \nonumber
 \end{eqnarray}
 Since $\vec{k}_1, \vec{k}_2, \vec{k}_3,$ and  $\vec{k}_4$ are dummy variables, we can switch $\vec{k}_4$ with $\vec{k}_2$ and $\vec{k}_3$ with $\vec{k}_1$, without changing the matrix element.  We thus finally have, for the second-order term,
 %\begin{widetext}
 \begin{eqnarray}
 %\fbox{$
 %\begin{array}{rcl}
\displaystyle\frac{d}{dt} \langle \hat{\rho}^{(2)}_{k,k',k'',k'''}\rangle &=& \displaystyle \frac{2\pi }{\hbar}(U_D\pm U_E)^2 \ \frac{1}{2}\sum_{{k}_2,{k}_3} \delta(E_{k}+E_{k_2}-E_{k_3}-E_{k_4}) \label{secondphase}\\
&&\hspace{.3cm} \times \langle \psi_i | \hat{\rho}^{(2)}_{{k,k',k'',k'''}}  \left[\pm
\hat{N}_{k_3}\hat{N}_{k_4}(1\pm \hat{N}_{k_2})% \right. \nonumber\\
%\\
%\hspace{1cm}\left.
-\hat{N}_{k_2}(1\pm \hat{N}_{k_3})(1\pm\hat{N}_{k_4}) \right]  | \psi_i \rangle   + \ldots .
%\hspace{1cm}+ N_{k'}N_{k_2}(1+N_{k_3})(1+N_{k_4})+N_{k_3}N_{k_4}(1+N_{k_2})(1+N_{k'}) \delta(E_{k'}+E_{k_2}-E_{k_3}-E_{k_4})\nonumber\\
%\hspace{1cm} +N_{k''}N_{k_2}(1+N_{k_3})(1+N_{k_4})+N_{k_3}N_{k_4}(1+N_{k_2})(1+N_{k''}) \delta(E_{k''}+E_{k_2}-E_{k_3}-E_{k_4})\nonumber\\
%\hspace{1cm} +N_{k'''}N_{k_2}(1+N_{k_3})(1+N_{k_4})+N_{k_3}N_{k_4}(1+N_{k_2})(1+N_{k'''}) \delta(E_{k'''}+E_{k_2}-E_{k_3}-E_{k_4}) \bigr].
%\end{array}
%$}
\nonumber
 \end{eqnarray}
 %\end{widetext}
where the $\ldots$ indicates that there are three more terms of exactly the same form, for $k'$, $k''$, and $k'''$ in $\hat{\rho}^{(2)}_{{k,k',k'',k'''}}$. 

Under the same assumption of no correlations that we used to factor (\ref{qboltz}), discussed in Section~\ref{sect.factor}, we can factorize (\ref{secondphase}) into products of expectation values. We then have
%\begin{widetext}
 \begin{eqnarray}
\frac{d}{dt} \langle \hat{\rho}^{(2)}_{k,k',k'',k'''}\rangle &=& \langle\hat{\rho}^{(2)}_{{k,k',k'',k'''}} \rangle \frac{2\pi }{\hbar}(U_D\pm U_E)^2 \ \frac{1}{2}\sum_{{k}_2,{k}_3} \delta(E_{k}+E_{k_2}-E_{k_3}-E_{k_4})\label{secondphase2}\\
&&\times  \left[\pm
\langle \hat{N}_{k_3}\rangle \langle \hat{N}_{k_4}\rangle (1\pm \langle \hat{N}_{k_2}\rangle)% \right. 
%\nonumber\\
%&&\left.
 -\langle \hat{N}_{k_2}\rangle(1\pm \langle \hat{N}_{k_3}\rangle)(1\pm\langle \hat{N}_{k_4}\rangle) \right]    + \ldots.
 \nonumber
 \end{eqnarray}
% \end{widetext}
The population factor $ %\left[
\langle\hat{N}_{k_3}\rangle\langle\hat{N}_{k_4}\rangle(1\pm \langle\hat{N}_{k_2}\rangle)%\right.$  $ \left.
-\langle\hat{N}_{k_2}\rangle(1\pm \langle \hat{N}_{k_3}\rangle)(1\pm \langle\hat{N}_{k_4}\rangle)$ %\right]$  
does not vanish in equilibrium, unlike the comparable term in the quantum Boltzmann equation, because it does not have the balancing factors $\langle\hat{N}_k\rangle$ and $(1\pm\langle \hat{N}_k\rangle)$.
For fermions, this rate is always negative, leading to dephasing, while for bosons, it can be positive if the net scattering rate into one of the $k$-states is positive. Net scattering into a state can occur, for example, when there is macroscopic coherence due to Bose condensation, which gives a macroscopic number of particles in a single state. Appendix C discusses the case of Bose condensation. At low density or high temperature, however, when $\langle N_k\rangle \ll 1$, the out-scattering term in (\ref{secondphase}) will dominate over the in-scattering term, since the in-scattering involves a product of two occupation numbers, while the out-scattering term has only a single occupation number. In other words, at low occupation per state, given a particle in state $k$, it is much more likely for that particle to scatter out of that state than for another to enter, in a given time interval. Therefore at low density the overall rate is always negative. %, even in equilibrium.
 
%At low density, the $\langle\hat{N}_{k_3}\rangle\langle\hat{N}_{k_4}\rangle$ term is small compared to the $\langle\hat{N}_{k_2}\rangle$ term, and so . 

%sum of four rates-- average to zero? can only be positive if all four are in region of net increase
  
%For bosons we must now look at the case when two of the operators in $\hat{V}_i$ do not commute with $\hat{\rho}^{(2)}_{{k,k',k'',k'''}}$. 

\subsection{Summary of Dephasing Calculation Results}
  
We have seen that on the one hand, that (\ref{firstphase}) implies that scattering does lead to deviation from a pure diagonal state, with an increase of the phase factor $\langle \hat{\rho}^{(2)}\rangle$ proportional to the net scattering rate. But in the case of fermions, or bosons at low density, that phase coherence can never build up; the result (\ref{secondphase2}) is of the form
\begin{equation}
 \displaystyle\frac{d}{dt} \langle \hat{\rho}^{(2)} \rangle = -\frac{\langle \hat{\rho}^{(2)} \rangle}{\tau},
\end{equation}
which gives exponential decay of the phase with a time constant $\tau$; the value of $\tau$ is equal to the time constant for out-scattering found from the quantum Boltzmann equation (\ref{avgboltz}). Moreover, in the low density limit, the phase build-up will always be negligible compared to the phase decay on average, because the build-up rate (\ref{firstphase}) is proportional to the square of the occupation number, while the decay term (\ref{secondphase2}) is linear in the occupation number at low density. 

Turning to the $\rho^{(4)}$ terms, which we found can also contribute to the evolution of $\langle \hat{N}_k\rangle$, it should not be hard to see that the same procedure applied to these phase factors will give the same behavior. In the above calculation for the dephasing, we could have just as easily inserted $\rho^{(4)}$, with the same result, but with eight scattering terms instead of four. In the same way, a first-order calculation like the one which gave us the result (\ref{firstphase}) for the accumulation of phase, presumably will also give us a term linear in $t$, but this phase accumulation will quickly be removed by the dephasing. 

We can therefore say in general terms that the natural evolution of the system makes the many-body states tend rapidly toward becoming diagonal states, which was the assumption used implicitly in the iterative solution of the quantum Boltzmann equation. In equilibrium, (\ref{firstphase}) implies that there is no build up of phase factors, and the states resolve to diagonal states, that is, states in which only $\langle \hat{N}_k\rangle$ is relevant. If the system is far from equilibrium, there will be a tiny buildup of phase in the superposition of states which is constantly being destroyed by the out-scattering processes.

Again, note that this entire calculation has been done for a closed system with energy conservation. No notions of measurement or collapse have been invoked. Rather, the deterministic evolution of the quantum many-body wave function in a large system gives rapid dephasing which leads to information loss. The strongest assumption we have made is that of Section~\ref{sect.factor}, that there is no correlation of the occupation numbers in different Fock states in an overall superposition. A similar assumption is made in Section~\ref{sect.dephase2}, that there is no correlation of phase and number in different states. 

Is the system reversible, then? Another way to put this question is whether we could choose some special initial state that would cause the system to evolve away from equilibrium and toward our initial state. The answer is yes, but our calculation gives an important qualification. To make the system evolve backwards, i.e., away from equilibrium, we would have to choose a special set of phase coherences in a superposition of states of the form (\ref{focksum}) that corresponded to the state of the system after some time evolution, set the system to have exactly these phase factors, and reverse the time evolution. But these phase factors tend exponentially quickly toward zero. Picking the right phase factors therefore is analogous to picking the proper initial state in a classical chaotic system to make it evolve toward a predetermined goal. As with predicting the motion of a classical chaotic system, time-reversing an interacting quantum many-body system involves exponential sensitivity to initial conditions (known as ``stiffness'') \cite{chaos}.

In all of this discussion, we have used the case of two-body, elastic interactions. It is not hard to see that the main results here do not depend on the exact form of the interactions. Any closed many-body system, even one in which particle number is not conserved, will evolve toward equilibrium and will have rapid dephasing of the information needed to reverse its evolution.

The above perturbative approach, in which we limit the calculation to the second order of perturbation theory, also assumes weakly interacting particles. If we did not assume weak interactions, then we could not treat the single-particle states as nearly eigenstates, as we have, and talk of particles scattering from one state to another. In other words, the whole particle picture would break down. In this case, what is typically done %for strongly interacting systems 
is to renormalize the system to define new eigenstates, which then become the new particles in the system (known as ``quasiparticles''), which have weak interactions between them. In this approach, a strongly interacting system can be mapped to a weakly interacting system with a new type of particle. (For example, plasmons are the new type of weakly interacting particle in a gas of strongly interacting electrons.) If this is not possible in some systems, then we cannot speak of equilibration of particles at all, because particles are no longer the proper description of the system. 

This calculation tells us something more than the classical Boltzmann scattering integral with the Stosszahlansatz. The quantum approach here says, first, that if we know that the system is in a purely diagonal state, then it will evolve deterministically toward equilibrium, and second, that the system naturally evolves toward diagonal states on time scales comparable to the scattering time. %and therefore it is natural to assume a diagonal state for the later states of the system as well.  
%The equivalent of the Stosszahlanzatz in the solution of the quantum Boltzmann equation is therefore to assume that once, in the first state, the system was not in a special choice of phases that leads the distribution of $\langle \hat{N}_k\rangle$ to move away from equilibrium. This is a very easy statistical assumption, since as we have seen, the off-diagonal factors decay exponentially rapidly toward zero---picking the right phases to reverse the evolution toward equilibrium would be many times harder than picking the right initial state of a chaotic system to cause it to evolve to a predetermined target. 
After starting in a nonequilibrium diagonal state, at a later time a system will not be in {\em exactly} a diagonal state; otherwise we could not time-reverse its state, which would violate the underlying time-reversibility. If we do not have exact knowledge of the initial state of the system, we cannot rule out the possibility that the initial state of the system was a state with all the phases exactly such that it will evolve away from equilibrium. 
As with the Stosszahlansatz in the classical H-theorem, we assume it is extremely unlikely to have been in such a state.  But since we know such states are exponentially suppressed in time, our choice of a diagonal state as the initial state in the absence of complete knowledge of the phases is not arbitrary, but naturally favored by the system itself. 

\vspace{-.25cm}
\section{Conclusions}

The picture that arises from quantum field theory is that the H-theorem and the Second Law arise, not from statistical arguments, but simply from the coupling of local energy to the whole vast universe.  If a system could truly be isolated perfectly, as a finite, energy-conserving system, then it would indeed undergo periodic motion forever. As soon as one allows any coupling to the rest of the universe, however, then there is damping which leads things to run down. This result has been deduced using exclusively the wave picture of quantum mechanics, without reference to wave-particle duality or the stochastic nature of identifying particle events with the wave function. No random numbers were used in the numerical calculations.  The H-theorem fundamentally comes from the nature of wave behavior in a system with an infinite number of degrees of freedom.  

%The equivalent of the Stosszahlansatz in the quantum mechanical derivation of the H-theorem is the assumption that the initial state is a diagonal (fully dephased) state; if it is then we can solve for the later evolution at all times using the quantum Boltzmann equation. This assumption can be viewed as a probabilistic one, that it would be highly unlikely to have been in an initial state which deviates strongly from a diagonal state. All states evolve toward diagonal states, with exponentially decreasing deviation from such states over time.

%append C-- relation of d-entropy and full entropy
As discussed in the introduction, work by Anatoli Polkovnikov \cite{polk} has shown that a ``diagonal entropy,'' or d-entropy, can be defined by summing only over diagonal states, and this entropy is extremely useful in calculations of thermodynamics properties in quantum systems. The present work shows that this choice of d-entropy is natural because the off-diagonal terms vanish quickly due to dephasing. The present work also bears some similarity to the work of Zurek \cite{zurek} and others which shows that quantum states can dephase to diagonal states due to interaction with an environment; in this work the ``environment'' is part of a closed, energy-conserving system, and all the dephasing is internal. 

Nuclear magnetic resonance spectroscopy provides an example of how this works.  
%Experimentally it is possible to measure only the spatially uniform single spin coherences.  
Suppose that a pick-up coil detects the global $x$-component of the magnetization (corresponding to a particular superposition of up and down spin) created by an initial $\pi/2$ tipping pulse.  Initially after the tipping pulse, the spin up and spin down parts of the wave function have the same spatial function.  However the spin up and spin down components of the particle take different trajectories and so eventually spatially separate from each other, entangling the spin with orbital motion.  The spin-up part of one packet can collide with the spin-down part of another packet, and these collisions spread the entanglement throughout the system just like the collisions we have considered in this paper.
%Even if we ignore the spin-lattice relaxation, the coupling of the spins with each other via dipole interactions moves this coherence into multi-spin correlations which are not experimentally detectable.  The equations of motion couple the $n$-spin operators to $n+1$-spin operators in a way that makes the superposition of up and down spins apparently decohere.  
The time evolution is completely unitary but produces multi-particle entangled states which look locally (i.e., in terms of individual spins) to be incoherent mixed states. %which have lost the transverse components of their magnetization.  
That the evolution is unitary and ultimately coherent can be seen in sophisticated echo techniques which can be used to narrow the resonance lines by orders of magnitude and recover the signal \cite{NMR}. 

Another example can be found in ultra-cold atomic gases which are quite isolated and essentially uncoupled from any heat baths.  Here the natural coherence is so high that it can actually be experimentally difficult to produce incoherent mixtures of different spin states.  In the case of cold fermions with short-range interactions, the Pauli principle renders these interactions ineffective except between opposite-spin particles.  Starting with all spin-up particles, application of a uniform $\pi/2$ pulse on all the spins does not turn on the interactions, even though both spin up and spin down are now present, because all the particles are in the same coherent superposition spin state.   However, in the presence of a spatial gradient in the Zeeman field, the spin state of the moving particles eventually becomes entangled in a complicated way with the spatial orbital state.  Despite the overall unitary time evolution, this leads to an effective decoherence of the spins, which turns on the interactions and leads to observable mean-field shifts in radio frequency spectroscopy experiments \cite{hulet,jin,ketterle}.

The above discussion in terms of off-diagonal terms of the generalized density matrix is made a little more complicated by the phenomenon of spontaneous phase coherence, which is seen experimentally in boson systems such as superfluids, superconductors, and lasers.  In such systems, as a system evolves toward equilibrium, a macroscopic fraction of the whole system can obtain a finite off-diagonal density matrix term (see Appendix C).  This fraction is known as the ``condensate.'' This leads to periodic behavior on long time scales, seen for example in the fascinating effects of persistent current and Josephson oscillations of superconductors. 

Furthermore, all classical wave states such as water waves on the ocean or radio waves can be viewed as condensates with macroscopic phase coherence. This is because all classical waves are technically macroscopic coherent states of bosons. (See Ref.~\cite{snokebook}, Section 4.4, and Ref.~\cite{mandel}, for discussion of coherent states, also known as Glauber states.) The symmetry breaking in this case does not come from a thermodynamic phase transition, %with a critical temperature, 
as occurs for superfluids and superconductors, nor even from amplification of small fluctuations, as in a laser, but from an external driving force;  thus classical waves are sometimes called ``driven condensates'' \cite{snoke-coherence}.  The external driving force breaks the symmetry and favors one definite phase over all others. If the driving force is removed, however, the interaction of many degrees of freedom eventually leads to dephasing. Classical waves typically can be sustained because the bosons of which they are comprised have very weak interactions; the phonons in sound waves have extremely weak interactions at low frequency (see Ref.~\cite{snokebook}, Sections 5.3 and 5.4), while photons in vacuum are essentially non-interacting. This allows coherent states to be sustained for long periods of time. Eventually, however, in the absence of an external driving force, they dephase and turn into a diagonal Planck equilibrium distribution, which we call heat.  

Condensates and classical waves have, to a high degree of approximation, reversible behavior. %Coupling of finite condensates to the rest of the universe means that coherent behavior in these systems always degrades eventually, though, and the overall evolution of the whole system is always toward equilibrium. 
No small system can ever be perfectly isolated from the rest of the universe, however. Because of %quantum mechanical 
tunneling, there will always be coupling to the infinite external universe, which will show up as damping in smaller systems. Thus all local systems dephase eventually due to radiation loss to the infinite space of the universe. This may occur extremely slowly, as in, for example, the cosmic background radiation which has not yet perfectly equilibrated over the eons, but the Second Law still holds.

This formulation of the Second Law has important cosmological implications. A prevalent model of the universe to account for the data of accelerating expansion \cite{cosmo} is that it is open, which means that it is infinite in spatial extent on any equal-time slice \cite{wheeler}, essentially an infinite ensemble of causally disconnected spaces. The possibility of an truly infinite ensemble can lead to bizarre implications if the standard, statistical version of the Second Law is used. For example, the fact that stars radiate energy rather than sucking energy in is due to the Second Law. If the universe is truly comprised of an infinite ensemble of locations, then on purely statistical grounds we would expect all finitely improbable events to occur, not just once, but an infinite number of times (though not necessarily observable to us). Thus we might expect, for example, that somewhere in the universe there are stars that suck in light rather than emit it (for any finite length of time). The quantum mechanical formulation presented here says that if stars are coupled to an infinite universe, then the Second Law is absolute, not just a likelihood, unless the beginning of the universe had very bizarre initial conditions of phase coherence. Any finite system coupled to the infinite background of the universe will have the same deterministic irreversibility. 

{\bf Acknowledgements}. We thank A. Duncan, D. Boyanovsky, and K. Sengupta for helpful discussions. This work has been partly supported by the National Science Foundation through grant DMR-010737. 

 \section*{Appendix A. Connection of the Boltzmann Equation to the H-Theorem and Entropy}

 \setcounter{equation}{0}
 \renewcommand{\theequation}{A.\arabic{equation}}

Here we derive the H-theorem from the quantum Boltzmann equation, following the standard approach based on the classical Boltzmann integral, e.g.  Ref.~\cite{tolman}.

For a quantum system, the total entropy is given by the von Neumann entropy %\cite{Bogo},
\begin{equation}
S_{vN} = -k_B {\sf Tr} (\tilde\rho \ln \tilde\rho),
\end{equation}
where $\tilde\rho$ is the full density matrix of the system. This is not the same as the single-particle density matrix defined in (\ref{singledens}), but instead is defined for the set of all Fock states. %which have the form $|n \rangle = | N_1,N_2,\ldots\rangle$, where $N_1, N_2, \ldots$ are the occupation numbers of the individual $k$-states. 
Using the general state (\ref{focksum}), we have $\tilde\rho_{nm} = \alpha_n^*\alpha_m$. Since the quantum mechanical evolution of a closed system involves only unitary operators, this entropy never changes. 

Polkovnikov \cite{polk} has suggested another entropy quantity, which can be called the {\em diagonal} entropy, %(d-entropy), 
\begin{equation}
S_d =  -k_B \sum_k \langle \hat{N}_k\rangle  \ln \langle \hat{N}_k \rangle.
\label{polkdiag}
\end{equation}
This corresponds to the standard thermodynamic entropy for an ensemble of classical particles. Like the concept of {\em entanglement} entropy \cite{entangle}, it involves a subset of the full quantum mechanical wave function information which is relevant to the macroscopic physical properties of the system. It also has relation to the concept of quantum coarse graining \cite{gell}, because it involves averaging over many states. 

The time derivative of the diagonal entropy (\ref{polkdiag}) is 
\begin{equation}
\frac{\partial S_d}{\partial t} = -k_B \sum_k \left(\frac{\partial \langle \hat{N}_k\rangle}{\partial t} \ln \langle \hat{N}_k\rangle + \frac{\partial \langle \hat{N}_k\rangle }{\partial t}\right).
\end{equation}
The second term vanishes due to number conservation. The first term is given by the quantum Boltzmann equation (\ref{avgboltz}), so that we have
\begin{eqnarray}
\frac{\partial S_d}{\partial t} &=& -  k_B C \sum_{k,k_1,k_2}  \ln \langle \hat{N}_k\rangle   \label{htheoremsum}  %\\
%&& \times 
\left[\langle \hat{N}_{{k}_1}\rangle \langle {\hat{N}}_{k_2}\rangle (1 \pm \langle \hat{N}_{k}\rangle)(1
\pm \langle \hat{N}_{k'}\rangle) \right.\nonumber\\&&\left.
\hspace{.5cm}
 -  \langle \hat{N}_{k}\rangle \langle \hat{N}_{k'}\rangle(1 \pm \langle
\hat{N}_{k_1}\rangle)(1 \pm \langle \hat{N}_{k_2}\rangle)\right]. 
\end{eqnarray}
where $C$ is a factor that contains the matrix element and conservation of energy factor.

If we pick four particular states $\vec{k}, \vec{k}', \vec{k}_1,$ and $\vec{k}_2$, then the total of all terms in the sum involving the same four states is 
\begin{eqnarray}
&& \left(\ln \langle \hat{N}_k\rangle+\ln \langle \hat{N}_{k'}\rangle-\ln \langle \hat{N}_{k_1}\rangle-\ln \langle \hat{N}_{k_2}\rangle \right)\nonumber\\
&&\times \left[\langle \hat{N}_{{k}_1}\rangle \langle {\hat{N}}_{k_2}\rangle (1 \pm \langle \hat{N}_{k}\rangle)(1
\pm \langle \hat{N}_{k'}\rangle) %\right.\nonumber\\
%&&\left.
%\hspace{.5cm}
-  \langle \hat{N}_{k}\rangle \langle \hat{N}_{k'}\rangle(1 \pm \langle
\hat{N}_{k_1}\rangle)(1 \pm \langle \hat{N}_{k_2}\rangle)\right] \nonumber\\
&=& \ln\left(\frac{\langle \hat{N}_k\rangle \langle \hat{N}_{k'}\rangle}{\langle \hat{N}_{k_1}\rangle \langle \hat{N}_{k_2}\rangle}\right)\label{H4sum} \\
&&\times \left[\langle \hat{N}_{{k}_1}\rangle \langle {\hat{N}}_{k_2}\rangle (1 \pm \langle \hat{N}_{k}\rangle)(1
\pm \langle \hat{N}_{k'}\rangle)% \right.\nonumber\\
%&&\left.
%\hspace{.5cm}
-  \langle \hat{N}_{k}\rangle \langle \hat{N}_{k'}\rangle(1 \pm \langle
\hat{N}_{k_1}\rangle)(1 \pm \langle \hat{N}_{k_2}\rangle)\right] . \nonumber
\end{eqnarray}
If the density is low, so that the $((1 \pm \langle \hat{N}_{k}\rangle)$ factors are negligible, this becomes
\begin{eqnarray}
 \ln\left(\frac{\langle \hat{N}_k\rangle \langle \hat{N}_{k'}\rangle}{\langle \hat{N}_{k_1}\rangle \langle \hat{N}_{k_2}\rangle}\right)  \left[\langle \hat{N}_{k_1}\rangle \langle {\hat{N}}_{k_2}\rangle -  \langle \hat{N}_{k}\rangle \langle \hat{N}_{k'}\rangle \right], 
\end{eqnarray}
which is always negative or zero for values of the occupation numbers greater than zero. Since the whole sum (\ref{htheoremsum}) consists of terms like this, the total sum is less than or equal to zero, and therefore $\partial S_d/\partial t > 0$. This is the standard form of the H-theorem. 

If the system obeys fermion statistics, the full sum (\ref{H4sum}) is also always negative or zero. But if the system obeys boson statistics which allow occupation numbers greater than unity, it can be made positive; for example, if $\langle \hat{N}_k\rangle = 10$, $\langle \hat{N}_{k'}\rangle = 0.1$, $\langle \hat{N}_{k_1}\rangle = 0.5$, and $\langle \hat{N}_{k_2}\rangle = 0.5$. Therefore, for some distribution functions $dS_d/dt < 0$. To avoid this problem, the proper {\em semiclassical} entropy quantity analogous to (\ref{polkdiag}) should be used. From standard quantum statistical mechanics \cite{thermobook}, this is
\begin{equation}
S_{sc} = -k_B\sum_k \left(\langle\hat{N}_k\rangle \ln \langle\hat{N}_k\rangle \mp (1\pm \langle\hat{N}_k\rangle)\ln(1\pm \langle\hat{N}_k\rangle)  \right),
\end{equation}
where the upper sign is for bosons and the lower sign is for fermions.
Assuming number conservation, the time derivative of this is
\begin{equation}
\frac{\partial S_{sc}}{\partial t} = -k_B\sum_k \frac{\partial \langle \hat{N}_k\rangle}{\partial t} \ln \left(\frac{ \langle \hat{N}_k\rangle}{  1\pm \langle \hat{N}_k\rangle} \right).
\end{equation}
Using the quantum Boltzmann equation for $\partial\langle\hat{N}_k\rangle/\partial t$ as above gives $\partial S_{sc}/\partial t > 0$ for all distributions.

%This is connected to the result (\ref{BECphase}) which shows that the coherent amplitude of a Bose condensate can be amplified rather than decay. In other words, the dephasing which we have seen leads to the Second Law can be reversed by the self-organization of a boson system at high density. As discussed in Appendix B, the increase of the phase amplitude ceases when the system reaches equilibrium; in the same way, $dS/dt$ does not keep increasing, but slows down to $dS/dt = 0$ in the calculation here when the system reaches equilibrium.

\section*{Appendix B. Calculation of the Off-Diagonal First-Order Terms}

 \setcounter{equation}{0}
 \renewcommand{\theequation}{B.\arabic{equation}}

As discussed in Section~\ref{sect.dephase1}, the first-order evolution of the $\hat\rho^{(2)}$ operator gives us terms of the following form:
%\begin{widetext}
\begin{eqnarray}
\langle [\hat\rho^{(2)}(t),\hat{V}_i(t') ]\rangle
&=& U\langle n' | e^{iH_0t}a^{\dagger}_k a^{\dagger}_{k'}a^{ }_{k''}a^{ }_{k'''}e^{-iH_0t}
e^{iH_0t'}a^{\dagger}_{k_4} a^{\dagger}_{k_3} a^{ }_{k_2}a^{ }_{k_1} e^{-iH_0t'} | n \rangle   \nonumber\\
&&   -    U\langle n' |
e^{iH_0t'}a^{\dagger}_{k_4} a^{\dagger}_{k_3} a^{ }_{k_2}a^{ }_{k_1} e^{-iH_0t'}e^{iH_0t}a^{\dagger}_k a^{\dagger}_{k'}a_{k''}a_{k'''}e^{-iH_0t}
| n \rangle. 
\end{eqnarray}
%\end{widetext}
We must treat a number of different possibilities: 1) one of the creation or destruction operators in $\hat{V}_i$ does not commute with one of the operators in $\hat\rho^{(2)}$; 2) two of the operators in $\hat{V}_i$ do not commute with $\hat\rho^{(2)}$; 3) three of the operators in $\hat{V}_i$ do not commute with $\hat\rho^{(2)}$; or 4) four of the operators in $\hat{V}_i$ do not commute with $\hat\rho^{(2)}$. We take these separately; each has several possibilities.

{\bf One non-commuting operator}. If there is one non-commuting operator in $\hat{V}_i$, we will have terms like the following:%xx this part is analogous to mean-field BEC phase rotation correction
%\begin{widetext}
\begin{eqnarray}
&&(a^{\dagger}_k  a^{ }_k  - a^{ }_k a^{\dagger}_k  ) a^{\dagger}_{k'}a^{ }_{k''}a^{ }_{k'''}a^{\dagger}_{k_4} a^{\dagger}_{k_3} a^{ }_{k_2} e^{i(E_{k}+E_{k'}- E_{k''}-E_{k'''})t/\hbar}e^{-i(E_{k}+E_{k_2}- E_{k_3}-E_{k_4})t'/\hbar} \nonumber\\
&=& -a^{\dagger}_{k'}a^{\dagger}_{k_4} a^{\dagger}_{k_3}a^{ }_{k''}a^{ }_{k'''} a^{ }_{k_2} e^{i(E_{k}+E_{k'}- E_{k''}-E_{k'''})t/\hbar}e^{-i(E_{k}+E_{k_2}- E_{k_3}-E_{k_4})t'/\hbar}.
\end{eqnarray}
%\end{widetext}
This is proportional to a third-order off-diagonal term $\hat\rho^{(3)}$ with six creation and destruction operators. We ignore these higher-order off-diagonal terms for the moment; in the spirit of perturbation theory, we can expand in a series of orders of off-diagonal terms, with the most significant being $\hat\rho^{(2)}$, assuming that as the degree of correlation increases, the lower the magnitude of the term. 

{\bf Two non-commuting operators}. There are three cases: a) two operators make up a number operator $\hat{N}_k$; b) two operators act on different $k$-states; or c) two operators act on the same state and are the same. 

a) In this case, we will have terms of like the following:
%\begin{widetext}
\begin{eqnarray}
&&(a^{\dagger}_k  \hat{N}_k  -\hat{N}_k a^{\dagger}_k  ) a^{\dagger}_{k'}a^{ }_{k''}a^{ }_{k'''} a^{\dagger}_{k_3} a^{ }_{k_2} e^{i(E_{k}+E_{k'}- E_{k''}-E_{k'''})t/\hbar}e^{-i(E_{k_2}- E_{k_3})t'/\hbar} \nonumber\\
&=& -a^{\dagger}_k a^{\dagger}_{k'}a^{ }_{k''}a^{ }_{k'''} a^{\dagger}_{k_3} a^{ }_{k_2} e^{i(E_{k}+E_{k'}- E_{k''}-E_{k'''})t/\hbar}e^{-i(E_{k_2}- E_{k_3})t'/\hbar} .
\end{eqnarray}
%\end{widetext}
If $\vec{k}_2 \ne \vec{k}_3$, we again have a higher-order $\hat\rho^{(3)}$ term. If $\vec{k}_2 = \vec{k}_3$, then this becomes
\begin{eqnarray}
 -a^{\dagger}_k a^{\dagger}_{k'}a^{ }_{k''}a^{ }_{k'''} \hat{N}_{k_2} e^{i(E_{k}+E_{k'}- E_{k''}-E_{k'''})t/\hbar} .
\end{eqnarray}
But in this case there is always also another term
\begin{eqnarray}
&&(a^{ }_{k'''}  \hat{N}_{k'''}  -\hat{N}_{k'''} a^{ }_{k'''}  )  a^{\dagger}_{k}a^{\dagger}_{k'} a^{ }_{k''}\hat{N}_{k_2} e^{i(E_{k}+E_{k'}- E_{k''}-E_{k'''})t/\hbar} \\
&& \hspace{1cm} = a^{\dagger}_k a^{\dagger}_{k'}a^{ }_{k''}a^{ }_{k'''} \hat{N}_{k_2} e^{i(E_{k}+E_{k'}- E_{k''}-E_{k'''})t/\hbar} ,\nonumber
\end{eqnarray}
which cancels the first term. This is true for both bosons and fermions.

b) In this case, we will have three types of terms. The first is:
\begin{eqnarray}
(a^{\dagger}_k  a^{ }_k a^{\dagger}_{k'}  a^{ }_{k'}  -  a^{ }_{k'} a^{\dagger}_{k'} a^{ }_k a^{\dagger}_k   ) a^{ }_{k''}a^{ }_{k'''} a^{\dagger}_{k_4} a^{\dagger}_{k_3} e^{i(E_{k}+E_{k'}- E_{k''}-E_{k'''})t/\hbar}e^{-i(E_{k}+E_{k'}- E_{k_3}-E_{k_4})t'/\hbar} .
%&=& (a^{\dagger}_k  a^{ }_k a^{\dagger}_{k'}  a^{ }_{k'}  -  (\pm a^{\dagger}_{k'}a^{ }_{k'} +1) ( \pm a^{\dagger}_ka^{ }_k+1   )) a^{ }_{k''}a^{ }_{k'''} a^{\dagger}_{k_4} a^{\dagger}_{k_3} t 
\nonumber\\
\end{eqnarray}
The time-dependent exponential factors in lowest order become just $t$ by the same type of calculation as in (\ref{firsttime}). So we have
\begin{equation}
 -(\pm \hat{N}_{k}\pm \hat{N}_{k'}+1)a^{ }_{k''}a^{ }_{k'''} a^{\dagger}_{k_4} a^{\dagger}_{k_3} t,
\end{equation}
where the $\pm$ signs give the + sign for bosons or $-$ sign for fermions.

The second type is:
\begin{eqnarray}
(a^{ }_{k''}  a^{\dagger}_{k''} a^{ }_{k'''}  a^{\dagger}_{k'''}  -  a^{\dagger}_{k''}  a^{ }_{k''} a^{\dagger}_{k'''}  a^{ }_{k'''}   ) a^{\dagger}_{k}a^{\dagger}_{k'} a^{ }_{k_2} a^{ }_{k_1}  t
%&=& ((\pm a^{\dagger}_{k''}a^{ }_{k''}  +1) (\pm a^{\dagger}_{k'''}a^{ }_{k'''}   +1) -  a^{\dagger}_{k''}  a^{ }_{k''} a^{\dagger}_{k'''}  a^{ }_{k'''}   ) a^{\dagger}_{k}a^{\dagger}_{k'} a^{ }_{k_1} a^{ }_{k_2} t 
= (\pm \hat{N}_{k''} \pm \hat{N}_{k'''}   +1) a^{\dagger}_{k}a^{\dagger}_{k'} a^{ }_{k_2}a^{ }_{k_1}  t,
 \nonumber\\
\end{eqnarray}
and the third type is:
\begin{eqnarray}
(a^{\dagger}_{k}  a^{ }_{k} a^{ }_{k'''}  a^{\dagger}_{k'''}  -  a^{ }_{k}a^{\dagger}_{k}  a^{\dagger}_{k'''}a^{ }_{k'''} ) a^{\dagger}_{k'}a^{ }_{k''} a^{\dagger}_{k_3} a^{ }_{k_2}  t 
%&=& (a^{\dagger}_{k}  a^{ }_{k}(\pm a^{\dagger}_{k'''} a^{ }_{k'''}+1)  - (\pm a^{\dagger}_{k} a^{ }_{k}+1)  a^{\dagger}_{k'''}a^{ }_{k'''} ) a^{\dagger}_{k'}a^{ }_{k''} a^{\dagger}_{k_3} a^{ }_{k_2}  t \nonumber\\
= ( \hat{N}_{k} -\hat{N}_{k'''} ) a^{\dagger}_{k'}a^{ }_{k''} a^{\dagger}_{k_3} a^{ }_{k_2}  t.
\end{eqnarray}
All of these depend on various $\rho^{(2)}$ factors.
% which are not the same as $\rho^{(2)}$. 
In general, they do not cancel, but one can expect that they average to zero when the sum over the free momenta is done.

c) These terms vanish for fermions and give higher order $\rho^{(3)}$ terms for bosons.

{\bf Three non-commuting operators}. There are two possibilities here: a) all three act on different $k$-states,  b) two act on one state, and the other acts on a different state.

a)  In this case, we will be left with factors that depend on the occupation numbers $N_k$ times a two-operator off-diagonal element of the form $\hat\rho_{k,k_2} = a^{\dagger}_ka^{ }_{k_2}$. If the system is in a diagonal state, this two-operator off-diagonal element cannot become nonzero through the two-body interaction we consider here. However, if there are nonzero $\rho^{(2)}$ terms, then the first-order evolution of $\hat\rho_{k,k_2}$ can be nonzero. 

b) This will lead to terms which depend on various $\rho^{(2)}$ factors. 

{\bf Four non-commuting operators}. This is the same as the case when all the states in $|n\rangle$ are restored, treated in Equation~(\ref{firstphase}).

\section*{Appendix C. Onset of Phase Coherence in a Bose-Einstein Condensate}

\setcounter{equation}{0}
\renewcommand{\theequation}{C.\arabic{equation}}

% connection to dephasing of Fock state of bosons
The rate of onset of phase coherence in a homogeneous condensate can be found be a simple extension of the second-order rate equations for phase already presented in Section~\ref{sect.dephase2}. Instead of looking at $\hat\rho^{(2)}_{k,k',k'',k'''} = a^{\dagger}_ka^{\dagger}_{k'}a^{ }_{k''}a^{ }_{k'''}$, we instead just insert $a_k$ in the same equations. %Since we calculated the evolution of $\hat\rho^{(2)}$ as four separate terms, each of which had only one of the operators in $\hat\rho^{(2)}$ not commuting with $\hat{V}_i$, 
In the case of a condensate we have just one term,
 \begin{eqnarray}
&&\frac{d}{dt} \langle a_k \rangle = \langle a_k \rangle \frac{2\pi }{\hbar}(U_D+ U_E)^2  
%&&\hspace{.5cm}\times 
\frac{1}{2}\sum_{{k}_2,{k}_3} \delta(E_{k}+E_{k_2}-E_{k_3}-E_{k_4})\label{BECphase} \\
&&\hspace{1.5cm}\times
 \left[
\langle \hat{N}_{k_3}\rangle \langle \hat{N}_{k_4}\rangle (1+ \langle \hat{N}_{k_2}\rangle) 
%\right.  \nonumber\\
%&&\hspace{1cm}\left. 
-\langle \hat{N}_{k_2}\rangle(1+ \langle \hat{N}_{k_3}\rangle)(1+\langle \hat{N}_{k_4}\rangle) \right]   .\nonumber
%&&\hspace{1.5cm}\times 
 \end{eqnarray}
 When there is net influx into the condensate state, any small coherence will be amplified, with expontential growth of the amplitude of the phase. The growth of the coherent amplitude will cease in equilibrium, when $N_k+1 \simeq N_k$. Then the condition $N_{k_3} {N}_{k_4} (1+ {N}_{k_2}) = 
{N}_{k_2}(1+ {N}_{k_3})(1+{N}_{k_4})$ is equivalent to $N_{k_3} {N}_{k_4} (1+ {N}_{k_2})(1+N_k) = 
N_k{N}_{k_2}(1+ {N}_{k_3})(1+{N}_{k_4})$, which is equal to zero in equilibrium.

The calculations for $\hat\rho^{(2)}$ can also be used to find the evolution of the phase of a condensate. The zero-order term (\ref{zerorot}) corresponds to 
\begin{eqnarray}
d \langle a_k \rangle = -i\frac{E_k}{\hbar}t  \langle a_k \rangle= -i\omega_k t \langle a_k \rangle.
\label{condrot}
\end{eqnarray}
In other words, the phase of the condensate rotates in time. The first-order term (\ref{firstexpans}) corresponds to
\begin{eqnarray}
&& d \langle \psi_i| a_k | \psi_i \rangle = \frac{1}{i\hbar}\int_0^t dt' \sum_{n,n'} \alpha_{n'}^*\alpha_n  \sum_{{k}_1,{k}_2,{k}_3}   \frac{U}{2}  e^{-iE_kt/\hbar} \\
&&\hspace{1.5cm}\times\left( \langle n' | a_{k}  e^{iH_0t'/\hbar}a^{\dagger}_{k_1}  a^{\dagger}_{k_2} a_{k_3}a_{k_4} e^{-iH_0t'\hbar} |n \rangle % \right. \nonumber \\
%&&\hspace{1.5cm}\left. 
- \langle n' |e^{iH_0t'/\hbar}a^{\dagger}_{k_1} a^{\dagger}_{k_2} a_{k_3}a_{k_4} e^{-iH_0t'/\hbar}  a_{k}  |n \rangle\right). \nonumber
\end{eqnarray}
The terms with $\vec{k}_1 = \vec{k}_4 = \vec{k}$ and $\vec{k}_2 = \vec{k}_3$, plus the equal terms with $\vec{k}_2=\vec{k}_3 = \vec{k}$ and $\vec{k}_1 = \vec{k}_4$, give us
\begin{eqnarray}
d \langle \psi_i| a_k | \psi_i \rangle &=&\frac{1}{i\hbar}\int_0^t dt' \sum_{n,n'} \alpha_{n'}^*\alpha_n \sum_{{k}_2}   {U}  e^{-iE_kt/\hbar} % \nonumber \\
%&&\hspace{1cm}\times 
\langle n' | (a^{ }_{k}a^{\dagger}_{k}-a^{\dagger}_{k}a^{ }_{k} ) \hat{N}_{k_2}a_{k}   |n \rangle. \nonumber \\
&\simeq & -i\frac{t}{\hbar} {U}  N_{\rm tot} \langle \psi_i| a_k | \psi_i \rangle,
\end{eqnarray}
where in the last step we took the limit $e^{-iE_kt/\hbar} \simeq 1$, for short times. This term gives us the first-order, mean-field renormalization of the condensate energy, which gives a shift of the frequency $\omega_k$ of rotation in (\ref{condrot}).

The time evolution which leads to onset of phase coherence in condensates has been an active topic of study over the past two decades \cite{bec}. The results here apply to a homogeneous gas in the thermodynamic limit, showing that spontaneous symmetry breaking occurs in the thermodynamic limit on short time scales, not only in the finite systems studied experimentally.
 
Several decades ago, Uhlenbeck \cite{uhlen} pondered the problem of using the assumption of coarse graining to justifying the Second Law of thermodynamics in connection with the macroscopic wavelike behavior of condensates. His intuition was basically correct: the assumptions which lead to irreversible behavior in normal systems break down in the case of condensates. The role of the coarse-graining assumption in the thinking of Uhlenbeck's day is here seen to be played by the fast-dephasing assumption.

We note that the above type of calculation applies to any Bose system, including ones without number conservation, such as phonons which a Planck distribution in equilibrium. Whenever occupation numbers exceed unity, there can be amplification of coherent phase fluctuations. It can be argued that this is the fundamental reason why macroscopic systems such as water waves 
are always in definite-amplitude states instead of Fock states. A macroscopic Fock state, which is a superposition of all different phases, i.e. a water wave which simultaneously has a crest and a trough, is a physically possible state.  The reason such is never seen is related to the spontaneous increase of phase coherence in boson systems.

%also-- entropy carried only by normal fluid 

\newpage
\end{document}